\documentclass[a4paper,oneside,11pt]{article}
\usepackage{graphicx,afterpage}
\usepackage{bm}\usepackage{soul}
\usepackage{empheq}\usepackage{mathtools}
\usepackage[top=0.6in, bottom=1in, left=1in, right=0.8in]{geometry}
\usepackage{mathrsfs,color}
\usepackage{amsfonts}\usepackage{multirow}
\usepackage{amssymb}\usepackage{caption,comment}
\usepackage{amsmath}\usepackage{float}
\usepackage{amssymb,amsthm}
\usepackage{graphicx,afterpage,cancel}

\usepackage[pdftex,colorlinks=true]{hyperref}
\hypersetup{CJKbookmarks,%
bookmarksnumbered,%
colorlinks,%
linkcolor=black,%
citecolor=black,%
plainpages=false,%
pdfstartview=FitH}
\numberwithin{equation}{section}
\numberwithin{figure}{section}

\newcommand\tabcaption{\def\@captype{table}\caption}

\definecolor{orange}{RGB}{255,127,0}

\def\d{{\, \rm d}}

\afterpage{\clearpage}

\def\d{{\, \rm d}}
\newcommand{\bfs}[1]{{\boldsymbol #1}}

\title{LEMDA: A Lagrangian-Eulerian Multiscale Data Assimilation Framework}
\author{Quanling Deng$^{1}$, Nan Chen$^{2*}$, Samuel N. Stechmann$^3$, and Jiuhua Hu$^{2}$  \\
{\footnotesize $^1$ School of Computing, Australian National University, Canberra, ACT 2601, Australia (Quanling.Deng@anu.edu.au)}\\
{\footnotesize $^2$ Department of Mathematics, University of Wisconsin–Madison, Madison, WI 53706, USA (chennan@math.wisc.edu)}\\
{\footnotesize $^3$ Department of Mathematics and Department of Atmospheric and Oceanic Sciences,} \\{\footnotesize University of Wisconsin–Madison, Madison, WI 53706, USA (Stechmann@wisc.edu)}\\
{\footnotesize $^*$ Corresponding author}}
\date{\today}
\begin{document}
\maketitle\tableofcontents
\abstract{Lagrangian trajectories are widely used as observations for recovering the underlying flow field via Lagrangian data assimilation (DA). However, the strong nonlinearity in the observational process and the high dimensionality of the problems often cause challenges in applying standard Lagrangian DA. In this paper, a Lagrangian-Eulerian multiscale DA (LEMDA) framework is developed. It starts with exploiting the Boltzmann kinetic description of the particle dynamics to derive a set of continuum equations, which characterize the statistical quantities of particle motions at fixed grids and serve as Eulerian observations. Despite the nonlinearity in the continuum equations and the processes of Lagrangian observations, the time evolutions of the posterior distribution from LEMDA can be written down using closed analytic formulae. This offers an exact and efficient way of carrying out DA, which avoids using ensemble approximations and the associated tunings. The analytically solvable properties also facilitate the derivation of an effective reduced-order Lagrangian DA scheme that further enhances computational efficiency. The Lagrangian DA within the framework has advantages when a moderate number of particles is used, while the Eulerian DA can effectively save computational costs when the number of particle observations becomes large. The Eulerian DA is also valuable when particles collide, such as using sea ice floe trajectories as observations. LEMDA naturally applies to multiscale turbulent flow fields, where the Eulerian DA recovers the large-scale structures, and the Lagrangian DA efficiently resolves the small-scale features in each grid cell via parallel computing. Numerical experiments demonstrate the skilful results of LEMDA and its two components.}

\section{Introduction}
Data assimilation (DA) seeks to optimally integrate different sources of information to improve the state estimation of a complex system \cite{kalnay2003atmospheric, law2015data, lahoz2010data, majda2012filtering, asch2016data}. Typically, the output from a numerical model is combined with observations to reduce the bias and the uncertainty in the estimated states, where observations are often noisy and sparse while the model usually contains model errors. DA plays a vital role in improving the state estimation at the initialization stage, facilitating the skilful prediction of chaotic or turbulent signals. DA is also widely used to recover the unobserved states or dynamically interpolate the partially observed time series. It is an essential tool for postprocessing the data in geophysics and climate science \cite{uppala2005era, kalnay1996ncep}.

Observational data can be categorized into two groups, depending on how they are measured. Eulerian observations refer to those estimated at fixed locations, while Lagrangian observations are often the drifters that follow a parcel of fluid's movement \cite{griffa2007lagrangian, blunden2019look, honnorat2009lagrangian, salman2008using, castellari2001prediction}. When these observations are combined with models for state estimation, the corresponding approaches are named Eulerian and Lagrangian DA, respectively. Eulerian observations are usually more straightforward to measure and convenient to use in DA, where the observational operator is linear in many practical applications. In contrast, Lagrangian DA is preferred in situations where traditional Eulerian observations are unavailable. It is a natural approach to recover the multiscale ocean velocity field, which is often not directly measured but can be inferred from the observed tracer trajectories \cite{apte2013impact, apte2008data, apte2008bayesian, ide2002lagrangian}. Some of the well-known Lagrangian data sets are the Global Drifter Program \cite{centurioni2017global}, which aims to estimate near-surface currents by tracking the surface drifters deployed throughout the global ocean, and the Argo program \cite{gould2004argo} that utilizes a fleet of robotic instruments drifting with the ocean currents for advancing the operational ocean DA. Lagrangian DA also provides a powerful tool to facilitate the recovery of the ocean eddies in the Arctic regions, where the sea ice floes play the role of the Lagrangian tracers \cite{mu2018arctic, chen2022efficient}. In addition to the ocean drifters, other types of Lagrangian tracers include trash or oil in the ocean \cite{van2012origin, garcia2022structured} and balloons collecting atmospheric data \cite{businger1996balloons}.

Despite the recent development of Lagrangian data sets, several significant challenges exist in applying Lagrangian DA to recover the underlying flow field. First, as the velocity field is a strongly nonlinear function of the displacement, the observational process of the Lagrangian DA is always highly nonlinear \cite{chen2014information, apte2008bayesian}. The strong nonlinearity in the observational process requires additional manipulations for applying standard ensemble-based DA methods \cite{sun2019lagrangian}. Second, the dimension of the coupled system for Lagrangian DA is usually quite large, which creates computational difficulties. The high dimensionality comes from both the underlying flow model and a large number of observed Lagrangian tracers. The latter is essential for recovering multiscale features of complex flow fields. Increasing the number of tracers by a significant amount is a straightforward way to allow the observations to collect information in the entire domain. However, such a strategy will significantly increase the computational cost of Lagrangian DA. Third, the trajectories of Lagrangian drifters may be affected by other external forcing sources in addition to the underlying flow field \cite{manucharyan2022subzero, potyondy2004bonded, cundall1979discrete}. For example, sea ice floe dynamics is complicated by collisions, where the contact forces between floes gradually dominate the drag force from the ocean when the number of the floes or, equivalently, the concentration increases. Modelling the exact instantaneous contact force is a highly challenging task, which substantially affects the accuracy of the state estimation from Lagrangian DA.

To overcome the mathematical and computational challenges in Lagrangian DA, a Lagrangian-Eulerian multiscale DA (LEMDA) framework is developed in this paper, where the directly available observations consist of only the Lagrangian trajectories. One crucial step in building such a framework is to use the Boltzmann equation as a kinetic-theory description of the particle dynamics to derive a set of continuum equations \cite{deng2023particle}, which characterize the statistical quantities of particle motions. It has been shown that using observed statistics for DA can resolve some challenging issues in applying standard trajectory observations \cite{bach2023filtering, farchi2021using}. Notably, this leads to an alternative consideration of the problem under the Eulerian framework, where the state variables describing the statistical features of the particles, such as the number of densities, the momentums, the angular momentums, etc., are defined at fixed grids. Since acquiring the observed Eulerian quantities requires a statistical average of the particle properties, these Eulerian observations are often calculated at coarse grids. Therefore, the focus on the Eulerian DA is to efficiently recover the large-scale features of the underlying flow field. In contrast, the observed individual particle trajectory allows the Lagrangian DA to infer a more refined flow field structure. One salient property of LEMDA is that, despite the nonlinearity in the continuum equations and the processes of Lagrangian observations, the time evolutions of the posterior distribution from both the Eulerian and Lagrangian DA can be written down using closed analytic formulae. This offers an exact and efficient way of carrying out DA, which avoids using ensemble methods and the associated tuning procedures.

In addition to significantly improving the computational efficiency by exploiting the closed analytic solution of the posterior distribution, the two components of LEMDA, namely the Eulerian and Lagrangian parts, respectively, have the following unique features. On the one hand, the new Eulerian DA can overcome at least two shortcomings in the standard Lagrangian DA. First, the computational cost of Lagrangian DA increases significantly as the number of observed trajectories increases. In contrast, the computational efficiency of the Eulerian DA relies only on the mesh size. With an increasing number of observed Lagrangian tracers, the averaged quantities of the Eulerian observations are estimated more accurately while the computational cost of the DA remains the same. Second, the statistical average in constructing the Eulerian observations automatically smooths out the interfering effect in recovering the underlying flow field due to particle collisions. In other words, the dominant forcing that drives the time evolution of the observed Eulerian statistical quantities comes solely from the flow field in such a situation. This unique feature provides an effective way for LEMDA to recover the large-scale flow structures in the presence of particle collisions, which is usually a challenging task for standard Lagrangian DA. On the other hand, rigorous mathematical theory can be applied to improve the computational efficiency of the nonlinear Lagrangian DA in the LEMDA framework under appropriate circumstances. The structure of the Lagrangian DA allows us to derive approximate schemes that remain analytically solvable for the DA solutions but further accelerate the computations. This includes predetermining the posterior covariance matrix for nearly incompressible flows when the number of tracers becomes large. Finally, the two DA approaches in the LEMDA framework can naturally be combined to develop a multiscale DA scheme that improves the state estimation of complex flow fields. The Eulerian DA aims to recover the large-scale features while the observed individual particle trajectory advances the Lagrangian DA to infer a more refined flow field structure within each grid cell. Notably, parallel computing can be implemented for applying the Lagrangian DA in different grid cells to significantly enhance computational efficiency.

The rest of the paper is organized as follows. Section \ref{Sec:Framework} describes the LEMDA framework. It includes the derivation of the continuum equation from the particle system and the approximate stochastic forecast models of the flow field. The section also contains the crucial analytic solution of the posterior distribution from DA. Section \ref{Sec:Quantification_Metrics} presents the path-wise and information metrics for assessing the DA skill. Section \ref{Sec:LagrangianDA} analyzes the performance of the Lagrangian DA and derives a reduced-order approximate Lagrangian DA scheme to significantly reduce the computational cost.  Section \ref{Sec:EulerianDA} studies the Eulerian DA and illustrates its advantages over the Lagrangian DA in the presence of particle collisions. Numerical experiments of LEMDA are presented in Section \ref{Sec:lemda}. The paper is concluded in Section \ref{Sec:Conclusion}.

\section{The LEMDA Framework}\label{Sec:Framework}
Consider a two-dimensional, doubly periodic domain $[0,2\pi)^2$. Assume there are $L$ Lagrangian tracers, whose displacements are denoted by $\bfs{x}_l = (x_l, y_l)^\mathtt{T}$ with $l = 1,\ldots,L$. Denote by $\bfs{v}_l = (u_l,v_l)^\mathtt{T}$ the corresponding tracer velocity. Further denote by $\bfs{u}=(\tilde{u},\tilde{v})^\mathtt{T}$ the underlying flow field that drives the motion of the Lagrangian tracers, which is also the target state variable to recover in DA.

LEMDA is a DA framework exploiting both the Lagrangian and Eulerian DAs. The Lagrangian DA uses the directly observed trajectories of the tracers. The Eulerian DA is based on a set of deduced continuum models via the Boltzmann description and the associated observed state variables. These quantities characterize the statistical properties of the particles described at fixed grids. Continuous-in-time observations are considered in this work and therefore the focus below is on continuous DA, which has been widely used in many theoretical studies \cite{azouani2014continuous, farhat2020data, bergemann2012ensemble, yang2013feedback, brocker2010variational, carlson2021sensitivity} and practical problems \cite{lean2021continuous, chen2022conditional, desamsetti2019efficient, rebholz2021simple}. The framework can be extended to its discrete-in-time DA analog.

\begin{figure}[ht]
    \hspace*{-0cm}\includegraphics[width=1.0\textwidth]{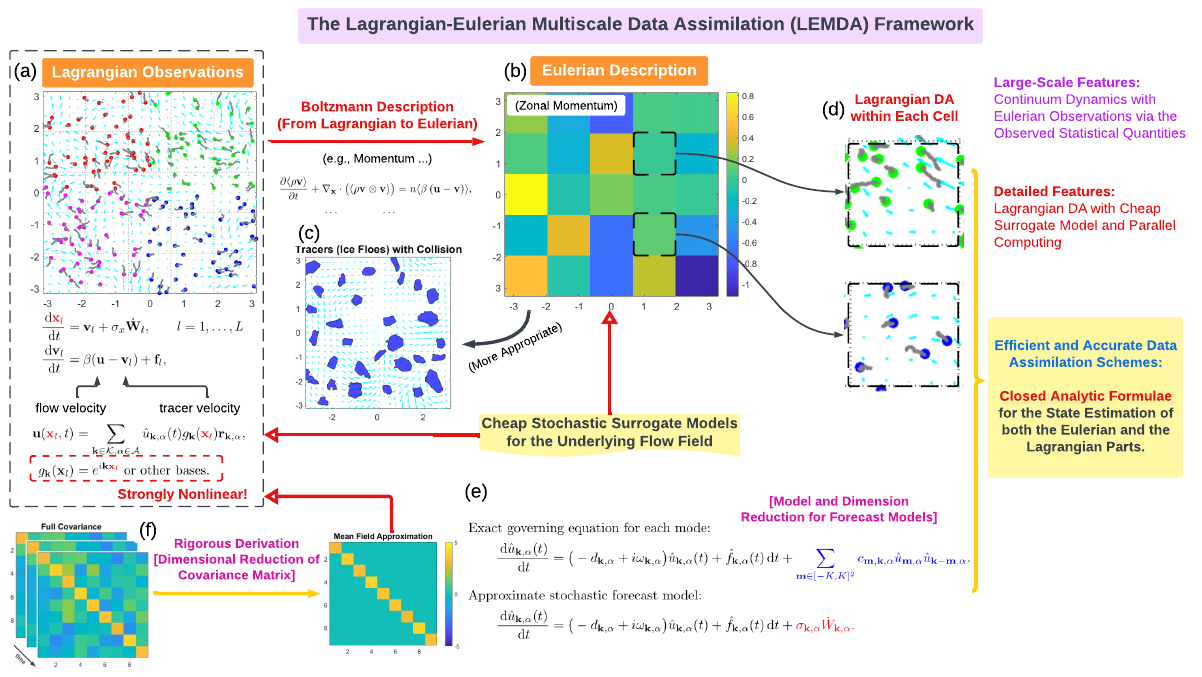}
    \caption{Overview of the LEMDA framework. Panel (a): the Lagrangian observations and Lagrangian DA (see Section \ref{Subsec:Lagrangian_Model}). Note the strong nonlinearity in the observational process of the Lagrangian DA. Panel (b): the Eulerian description of particle statistics, which can be derived from the Boltzmann description of the particle system (see Section \ref{Subsec:Eulerian_Model}). Panel (c): applying the Eulerian DA to the cases where the observed tracers are the ice floes with collisions (see Section \ref{Subsec:Collisions}). Panel (d): applying the Lagrangian DA to each grid cell to recover refined features that are missed by the large-scale Eulerian DA in the multiscale DA framework. Panel (e): the development of cheap stochastic surrogate models for the underlying flow field, which is a crucial part of allowing the analytic solvers of both the Eulerian and the Lagrangian DAs (see Section \ref{Subsec:Approximate_Models}). Panel (f): rigorous derivation of reduced-order DA schemes for Lagrangian DA (see Section \ref{Subsec:Reduced_LDA}). The closed analytic solutions for the posterior distributions of LEMDA are shown in Section \ref{Subsec:Closed_Formulae}.   }
    \label{LEMDA_Overview}
\end{figure}

Figure \ref{LEMDA_Overview} includes an overview of the LEMDA framework. The details of each component will be presented in the remaining subsections.

\subsection{The Lagrangian particle model}\label{Subsec:Lagrangian_Model}
The Lagrangian particle model describes the motion of each tracer:
\begin{subequations}\label{Lagrangian_Tracer_Model}
\begin{align}
  \frac{\d { \bfs{x}_l}}{\d t} &= \bfs{v}_l + \sigma_x\dot{\bfs{W}}_l, \label{Lagrangian_Tracer_Model_x}\\
  \frac{\d \bfs{v}_l}{\d t} &= \beta(\bfs{u}-\bfs{v}_l) + \bfs{f}_l,\label{Lagrangian_Tracer_Model_v}
\end{align}
\end{subequations}
where $\dot{\bfs{W}}_l$ in \eqref{Lagrangian_Tracer_Model_x} is a Gaussian white noise accounting for the measurement noise or the contribution from the unresolved scales of the flow field, and $\sigma_x$ is the noise coefficient. In \eqref{Lagrangian_Tracer_Model_v}, $\beta$ is the drag coefficient, where the drag force provides the basic mechanism for the underlying flow field to drive the tracer velocity. For a finite value of $\beta$, the response of the tracer velocity to the ocean has a delayed effect. The drag coefficient is usually proportional to the inverse of the mass of the particle. In many applications, for example, when sea ice floes represent the particles, the drag coefficient is around order $O(1)$ when the equations are written in the standard non-dimensional form, as is used in this work \cite{chen2021lagrangian}. Only a linear drag is used in \eqref{Lagrangian_Tracer_Model} for simplicity. The equation can be easily modified by utilizing other types of drag force \cite{damsgaard2018application}. The forcing term $\bfs{f}_l$ represents the other external forces, such as the one due to the particle collision (e.g., when the trajectories of the ice floes are used as Lagrangian observations). The forcing $\bfs{f}_l$ is typically a nonlinear function of $\bfs{x}_l$ and can be a function of $\bfs{v}_l$ and $\bfs{u}$ as well \cite{cundall1979discrete, chen2021lagrangian}. If $\bfs{f}_l=\bfs{0}$, then in the limit of $\beta\to\infty$, the flow velocity $\bfs{u}$ and the tracer velocity $\bfs{v}_l$ are completely synchronized, which has also been used in many theoretical analysis and idealized tests \cite{apte2013impact, apte2008data, apte2008bayesian, ide2002lagrangian}.

\subsection{The continuum model for describing Eulerian quantities}\label{Subsec:Eulerian_Model}
The Boltzmann equation was originally proposed to describe the statistical behaviour of a thermodynamic system \cite{cercignani1988boltzmann}. The Lagrangian tracers are analogs to the molecules in thermodynamics. Mathematically, by assuming the tracer number goes to infinity, the Boltzmann-type equations can be naturally derived.

The phase space of the Boltzmann equation is given by the set of all possible positions $\bfs{x}$ and velocity $\bfs{v}$. The differential volume element is written as
\begin{equation}\label{Differential_Volume_Element}
  \d E = \d \bfs{x} \d \bfs{v}.
\end{equation}
In the limit $L\to\infty$, the general Boltzmann equation describing the statistical behaviour of the particle motion is given by \cite{harris2004introduction}
\begin{equation} \label{General_Boltzmann}
    \frac{\partial g}{\partial t}
    + \nabla_{\bfs{x}} \cdot \Big(g \frac{\d \bfs{x}}{\d t} \Big)
    + \nabla_{\bfs{v}} \cdot \Big(g \frac{\d \bfs{v}}{\d t} \Big)
    = C(g),
\end{equation}
where $g$ is a density function of the space $(t, \bfs{x}, \bfs{v})$. The term $C(g) = \Big( \frac{\partial g}{\partial t}\Big)_{\text{coll}}$ represents the particle interaction, such as the collision in the ice floe case, which satisfies
\begin{equation} \label{eq:sumcf0}
    \int C(g) \d E = 0.
\end{equation}
It is worth noting that extra variables can be added to the phase space. For example, angular displacement and angular velocity are additional crucial state variables when the shape of particles is considered, such as the ice floes. The Boltzmann equation in \eqref{General_Boltzmann} already has five dimensions. The dimension will be further increased with additional variables. Solving directly the Boltzmann equation raises numerical challenges.

To allow practical numerical simulation of the particle statistics, the equations of the so-called statistical moments are derived to reduce the dimension of the system while maintaining the main large-scale dynamical features. To this end, integrating the Boltzmann equation \eqref{General_Boltzmann} over the velocity space leads to the continuity equation for the number densities $\rho$:
\begin{equation}\label{eq:numdensity}
    \frac{\partial \rho }{\partial t} + \nabla_{\bfs{x}} \cdot \big(\rho \langle \bfs{v} \rangle \big) = 0,
\end{equation}
where $\langle \bfs{v} \rangle$ is the particle velocities at fixed locations. Practically, $\langle \bfs{v} \rangle$ can be regarded as the averaged velocity of the particles appearing in a grid cell. Similarly, the momentum equations can be derived
\begin{equation}\label{eq:mv}
    \frac{\partial \langle \rho \bfs{v}  \rangle }{\partial t}
    + \nabla_{\bfs{x}} \cdot \big( \langle \rho \bfs{v} \otimes \bfs{v} \rangle \big)
     = n \langle \beta\left(\bfs{u} -\bfs{v}\right) \rangle,
\end{equation}
where $\otimes$ is the outer product and $n$ is the particle number density.

The flow velocity $\bfs{u}$ appears on the right-hand side of \eqref{eq:mv}, allowing the equation to be used as an observational process for the Eulerian DA. The observation is the momentum $\langle \rho \bfs{v}  \rangle$, which can be computed by exploiting the observed Lagrangian tracers. Depending on the applications, other statistical moment equations, such as those describing the angular momentums, can be derived \cite{deng2023particle}.

In this paper, the two components of the momentum in \eqref{eq:mv} will be used as observations for the Eulerian DA.

\subsection{Stochastic approximate forecast model for the flow field}\label{Subsec:Approximate_Models}
DA involves running the forecast model to obtain the so-called prior distribution \cite{evensen2000ensemble}. The forecast model of the underlying flow field is typically complicated and expensive to run. Let alone using an ensemble of the forecast trajectories in the ensemble DA, which often leads to a significant computational challenge in practice.

In this subsection, a general framework for approximating the forecast model that characterizes the underlying turbulent flow field is developed. With appropriate stochastic approximations, the forecast prior distribution remains similar to that using the original model, but the computational cost is reduced significantly. Notably, the structure of the forecast model advances the use of closed analytic formulae to solve the DA solution, which avoids using expensive ensemble methods and will be discussed in the subsequent sections.

Consider the following spectral representation of the underlying flow velocity field \cite{chen2015noisy, majda2003introduction},
\begin{equation}\label{Ocean_Velocity}
  \bfs{u}(\bfs{x},t) = \sum_{\bfs{k}\in\mathcal{K},\alpha\in\mathcal{A}}\hat{u}_{\bfs{k},\alpha}(t)e^{i\bfs{k}\bfs{x}}\bfs{r}_{\bfs{k},\alpha},
\end{equation}
where $\bfs{x}=(x,y)^\mathtt{T}$ is the two-dimensional coordinate and the flow field is given by a finite summation of Fourier modes. The index $\bfs{k}=(k_1,k_2)^\mathtt{T}$ is the wavenumber, and the index $\alpha$ represents different types of waves, including, for example, the gravity modes and the geophysically balanced modes in the study of many geophysical flows. The set $\mathcal{K}$ usually consists of all the wavenumbers that satisfy $-K_{\mbox{max}}\leq k_1, k_2\leq K_{\mbox{max}}$ with $K_{\mbox{max}}$ being an integer that is pre-determined. The vector $\bfs{r}_{\bfs{k},\alpha}$ is the eigenvector, which links the two components of velocity fields, namely $\tilde{u}$ and $\tilde{v}$. As the left hand side of \eqref{Ocean_Velocity} is evaluated at physical space, the coefficients $\hat{u}_{\bfs{k},\alpha}$ and $\hat{u}_{-\bfs{k},\alpha}$ for all $\bfs{k}$ are complex conjugates. So do the eigenvectors $\bfs{r}_{\bfs{k},\alpha}$ and $\bfs{r}_{-\bfs{k},\alpha}$. Note that the Fourier basis functions are adopted here to simplify the description of the framework. Different basis functions and boundary conditions can be utilized in \eqref{Ocean_Velocity} for various applications in practice.

Stochastic models are used to forecast the time evolution of each Fourier coefficient $\hat{u}_{\bfs{k},\alpha}$ in \eqref{Ocean_Velocity} in the forecast stage of DA, which is a much more computationally efficient way to mimic the observed turbulent flows generated from a complicated partial differential equation (PDE) system. Although stochastic models will not reproduce the exact single forecast trajectory from the original system, the forecast prior distribution, as is required in DA, from appropriate stochastic models can be similar to that from the more complicated model. To this end, the stochastic model is often calibrated by matching several key statistics in the observed time series of $\hat{u}_{\bfs{k},\alpha}$. Among different stochastic models, the linear stochastic model, namely the complex Ornstein-Uhlenbeck (OU) process \cite{gardiner1985handbook}, is a widely used choice:
\begin{equation}\label{OU_process}
  \frac{\d\hat{u}_{\bfs{k},\alpha}}{\d t} = (- d_{\bfs{k},\alpha} + i\omega_{\bfs{k},\alpha}) \hat{u}_{\bfs{k},\alpha} + f_{\bfs{k},\alpha}(t) + \sigma_{\bfs{k},\alpha}\dot{W}_{\bfs{k},\alpha},
\end{equation}
where $d_{\bfs{k},\alpha}, \omega_{\bfs{k},\alpha}$ and $f_{\bfs{k},\alpha}(t)$ are damping, phase and deterministic forcing, $\sigma_{\bfs{k},\alpha}$ is the noise coefficient and $\dot{W}_{\bfs{k},\alpha}$ is a white noise. The constants $d_{\bfs{k},\alpha}$, $\omega_{\bfs{k},\alpha}$ and $\sigma_{\bfs{k},\alpha}$ are real-valued while the forcings are complex. The stochastic noise in the linear stochastic model is utilized to effectively parameterize the nonlinear deterministic time evolution of chaotic or turbulent dynamics \cite{majda2016introduction, farrell1993stochastic, berner2017stochastic, branicki2018accuracy, majda2018model, li2020predictability, harlim2008filtering, kang2012filtering}.
The four parameters $f_{\bf k,\alpha}$, $d_{\bf k,\alpha}$, $\omega_{\bf k,\alpha}$ and $\sigma_{\bf k,\alpha}$ in the complex-valued linear stochastic model \eqref{OU_process} can be uniquely determined by utilizing the four statistics of $\hat{u}_{\bf k,\alpha}$: the mean $m_{\bf k,\alpha}$, the variance $E_{\bf k,\alpha}$, and the real and imaginary parts of the decorrelation time $T_{\bf k,\alpha}- i\theta_{\bf k,\alpha}$ \cite{majda2012filtering},
\begin{equation}\label{eq:matching_statistics}
f_{\bf k,\alpha}=\frac{(T_{\bf k,\alpha}-i\theta_{\bf k,\alpha})m_{\bf k,\alpha}}{T_{\bf k,\alpha}^2+\theta_{\bf k,\alpha}^2},~
 d_{\bf k,\alpha}=\frac{T_{\bf k,\alpha}}{T_{\bf k,\alpha}^2+\theta_{\bf k,\alpha}^2},~
\omega_{\bf k,\alpha}=\frac{\theta_{\bf k,\alpha}}{T_{\bf k,\alpha}^2+\theta^2_{\bf k,\alpha}},~
\sigma_{\bf k,\alpha}=\sqrt{\frac{2E_{\bf k,\alpha}T_{\bf k,\alpha}}{T_{\bf k,\alpha}^2+\theta_{\bf k,\alpha}^2}}.
\end{equation}
There are two scenarios in practice. First, if the true forecast model is known, then the above four statistics can be calculated based on a long simulation from the model. With these statistics in hand, the relationships in \eqref{eq:matching_statistics} are utilized to determine the parameters in the linear stochastic model \eqref{OU_process} associated with each spectral mode. Second, if the true forecast model is unknown, then an iterative method can be adopted to alternate between estimating the parameters in the linear stochastic model and the DA. This is achieved by starting with a random guess of the parameters to recover the flow field driven by the set of linear stochastic models. It is followed by updating the parameter values from the recovered time series of each spectral mode. Repeat this procedure until the convergence. See \cite{chen2023uncertainty} for details.

The mathematical framework of modelling the random flow field with linear stochastic models characterizing the time series of spectral modes in \eqref{Ocean_Velocity}--\eqref{OU_process} has been widely used to describe various turbulent flow fields, including the rotating shallow water equation \cite{chen2015noisy} and the quasi-geostrophic equation \cite{chen2023stochastic, covington2022bridging}. It has been adopted as an effective surrogate forecast model in DA to approximate the Navier-Stokes equations \cite{branicki2018accuracy}, moisture-coupled tropical waves \cite{harlim2013test} and a nonlinear topographic barotropic model \cite{chen2023uncertainty}. Quantifying the uncertainty using the linear stochastic model as a surrogate model in the statistical forecast and filtering can be found in \cite{branicki2013non, chen2023uncertainty, chen2016model}.

In this paper, the linear stochastic models \eqref{OU_process} will be used to generate the true signal. They will also be served the forecast model of the flow field for the DA. In practice, the true system can be highly nonlinear while the calibrated linear stochastic models \eqref{OU_process} can remain as the forecast model, as discussed above. Note that the observational processes in both Eulerian and Lagrangian DA are nevertheless highly nonlinear. Therefore, the overall DA remains nonlinear.

The strong nonlinearity in DA can be seen as follows. The Eulerian DA consists of the observational process of the statistical moments, i.e., the momentum \eqref{eq:mv} in this work, and the underlying flow field driven by a set of stochastic equations \eqref{OU_process}. They are linked by the spectral representation of the flow field \eqref{Ocean_Velocity}. The coupled DA system is nonlinear due to the terms on the left-hand side of \eqref{eq:mv}. Similarly, the Lagrangian observations are driven by the particle system \eqref{Lagrangian_Tracer_Model}. Notably, once the flow velocity $\bfs{u}$ in \eqref{Lagrangian_Tracer_Model} is written explicitly by \eqref{Ocean_Velocity}, the state variable $\bfs{x}$ of the Lagrangian observations appear in the exponential function on the right-hand side of the governing equation. Therefore, both the Eulerian and Lagrangian DAs are strongly nonlinear.
\subsection{The LEMDA framework with closed analytic solution}\label{Subsec:Closed_Formulae}

Despite the strong nonlinearity, the posterior solutions of both the Eulerian and the Lagrangian DAs can be written down using closed analytic formulae. This crucial feature can be seen by noticing that the two DA systems can both be written in the following abstract form:
\begin{subequations}\label{CGNS}
\begin{align}
\frac{\d \bfs{Z}(t)}{\d t} &= \bfs{F}_\bfs{Z}(\bfs{Z}, t) + \bfs{A}(\bfs{Z}, t) \bfs{U}(t) + \bfs{\Sigma}_\bfs{Z} \dot{\bfs{W}}_\bfs{Z}(t),\label{CGNS_X}\\
\frac{\d \bfs{U}(t)}{\d t} &=  \bfs{F}_\bfs{U}(\bfs{Z}, t) + \bfs{\Lambda} \bfs{U}(t)  + \bfs{\Sigma}_\bfs{U} \dot{\bfs{W}}_\bfs{U}(t),\label{CGNS_U}
\end{align}
\end{subequations}
where $\bfs{F}_\bfs{Z}(\bfs{Z}, t)$, $\bfs{F}_\bfs{U}(\bfs{Z}, t)$ and $\bfs{A}(\bfs{Z}, t)$ are nonlinear functions of $\bfs{Z}$, and $\bfs{\Sigma}_\bfs{Z} \dot{\bfs{W}}_\bfs{Z}(t)$ and $\bfs{\Sigma}_\bfs{U} \dot{\bfs{W}}_\bfs{U}(t)$ are Gaussian white noises.
In the Lagrangian DA, $\bfs{Z}$ is the collection of the $L$ Lagrangian tracer trajectories. Therefore, \eqref{CGNS_X} contains the $2L$ equations in the form of  \eqref{Lagrangian_Tracer_Model_x}, describing $x_l$ and $y_l$ for $l=1,\ldots,L$. The state variable $\bfs{U}$ is the collection of the spectral modes $\hat{u}_{\bfs{k},\alpha}$ of the flow field driven by \eqref{OU_process} and the velocity $\bfs{v}_l$ of different tracers \eqref{Lagrangian_Tracer_Model_v}. In the Eulerian DA, $\bfs{Z}$ becomes the observed statistical moments, e.g., the momentums $\langle \rho \bfs{v}\rangle$, of the tracers. In other words, the governing equation of $\bfs{Z}$ is the discrete form of \eqref{eq:mv} at discrete mesh grids, where $N_x$ and $N_y$ grids are utilized in the $x$ and the $y$ directions, respectively. Additional random noise is added to the resulting equations to compensate for the discretization error and observational error. The latter is mainly due to computing these statistics based on a finite number of particles. See the Supporting Information for details. The state variable $\bfs{U}$ corresponds to the underlying flow velocity driven by \eqref{OU_process}.

The DA problems in \eqref{CGNS} are highly nonlinear in terms of the joint variables $(\bfs{Z},\bfs{U})$. Nevertheless, conditioned on the observational variable $\bfs{Z}$, the equations of the unobserved states $\bfs{U}$ that contain the underlying flow field are linear. Notably, the associated conditional distribution of $\bfs{U}$ is precisely the DA solution. Due to the conditional linearity, given one realization of the observed variable $\bfs{Z}(s\leq t)$, the filtering posterior distribution $p(\bfs{U}(t)|\bfs{Z}(s\leq t))$ of the DA problem in \eqref{CGNS} is conditionally Gaussian, where the time evolutions of the conditional mean $\bfs\mu$ and the conditional covariance $\bf R$ have the following closed analytic formulae \cite{liptser2013statistics, chen2018conditional}
\begin{subequations}\label{eq:filter}
\begin{align}
\frac{\d\bfs{\mu}}{\d t} &= \left(\bfs{F}_\bfs{U} + \bfs{\Lambda} \bfs{\mu}\right)  + \bfs{R}\bfs{A}^\ast(\bfs{\Sigma}_\bfs{Z}\bfs{\Sigma}_\bfs{Z}^\ast)^{-1}\left(\frac{\d \bfs{Z}}{\d t} - (\bfs{F}_\bfs{Z}+\bfs{A}\bfs{\mu}) \right),\label{eq:filter_mu}\\
\frac{\d\bfs{R}}{\d t} &= \bfs{\Lambda}\bfs{R} + \bfs{R}\bfs{\Lambda}^\ast + \bfs{\Sigma}_\bfs{U}\bfs{\Sigma}_\bfs{U}^\ast - \bfs{R}\bfs{A}^\ast(\bfs{\Sigma}_\bfs{Z}\bfs{\Sigma}_\bfs{Z}^\ast)^{-1}\bfs{A}\bfs{R},\label{eq:filter_R}
\end{align}
\end{subequations}
with $\cdot^*$ being the complex conjugate transpose. The continuous filtering solution in \eqref{eq:filter} can be regarded as a generalized nonlinear version of the Kalman-Bucy filter \cite{kalman1961new, bucy2005filtering, jazwinski2007stochastic}. Nevertheless, the nonlinear nature of the coupled system leads to a random Riccati equation for solving the covariance matrix. In addition, the gain matrix $\bfs{R}\bfs{A}^\ast(\bfs{\Sigma}_\bfs{Z}\bfs{\Sigma}_\bfs{Z}^\ast)^{-1}$ is time-dependent due to the nonlinearity in the observational process. These properties distinguish the formulae in \eqref{eq:filter} from the standard Kalman-Bucy filter and facilitate the filtering solution to better capture turbulent features triggered by strong nonlinearity.

Besides filtering, analytic formulae are available for the posterior distribution of the nonlinear smoother \cite{chen2020learning}, which is widely used to dynamically interpolate the states and obtain the reanalysis data \cite{sarkka2023bayesian, einicke2012smoothing}. This paper focuses on the filtering, but LEMDA can be easily extended to seek the nonlinear smoother solution.

With these analytic formulae for the posterior estimates, ensemble methods are no longer needed in solving the DA problems. Therefore, localization, covariance inflation, and other tuning procedures can be avoided. This is one of the major advantages of LEMDA, which facilitates exact, accurate, and robust solutions. In addition, with the analytic formulae, computational efficiency is significantly improved.


\section{Assessing the Skill of DA}\label{Sec:Quantification_Metrics}
\subsection{Path-wise measurements for assessing the skill in the posterior mean estimate}
Two path-wise measurements are utilized to quantify the skill of DA. They are the root-mean-square error (RMSE) and the pattern correlation (Corr)\cite{hyndman2006another}. These two measurements compare the posterior mean and the truth. Given two vectors $\bfs{a}=(a_1,a_2,\ldots,a_I)$ (posterior mean) and $\bfs{b}=(b_1,b_2,\ldots,b_I)$ (truth). They are defined as:
\begin{subequations}\label{Skill_Scores}
\begin{align}
\mbox{RMSE} &= \left(\sqrt{\frac{\sum_{i=1}^I(a_i-b_i)^2}{I}}\right)/\mbox{std}(\bfs{b}),\label{Skill_Scores_RMSE}\\
\mbox{Corr} &= \frac{\sum_{i=1}^I\Big(\big(a_i-\mbox{mean}{(\bfs{a})}\big)\big(b_i-\mbox{mean}{(\bfs{b})}\big)\Big)}
{\sqrt{\sum_{i=1}^I\big(a_i-\mbox{mean}{({\bfs{a}})}\big)^2}\sqrt{\sum_{i=1}^I\big(b_i-\mbox{mean}{(\bfs{b})}\big)^2}},\label{Skill_Scores_Corr}
\end{align}
\end{subequations}
where mean$(\bfs{a})$ and mean$(\bfs{b})$ are the mean value of the vectors $\bfs{a}$ and $\bfs{b}$, respectively, and std$(\bfs{b})$ is the standard deviation of $\bfs{b}$. The RMSE is non-negative. A more skilful posterior mean estimate corresponds to a smaller RMSE. Note that the RMSE defined in \eqref{Skill_Scores_RMSE} has already been normalized by the standard deviation of the truth. Therefore, when RMSE is above $1$, the error in the posterior mean estimate is larger than the mean of the truth. The pattern correlation gives a number between $-1$ and $1$. A larger value of the pattern correlation corresponds to a more similar profile between the two vectors. Usually, a skilful estimate has a pattern correlation larger than $0.5$. In this paper, the DA solution in spectral space is first projected to the two-dimensional physical space. The two-dimensional data field is reshaped to be stored in a vector. Then, these skill scores are applied to the two vectors of the truth and the posterior mean estimate at each time instant. Either the time series of the skill scores or the time average value will be shown in the following sections.

\subsection{An information measurement for assessing the skill in the posterior distribution}
Another useful measurement aims to quantify the uncertainty reduction in DA, which uses the information from the entire posterior distribution. The total uncertainty in DA can be quantified by computing the difference between the posterior and the prior distributions. Here, the prior distribution is the statistical equilibrium state of the forecast model, which does not involve additional information from observations.
The relative entropy, which is an information criterion, can be utilized to assess the uncertainty reduction (or information gain) in the posterior distribution $p(t)$ compared with the prior one $p_{eq}$ \cite{majda2005information, kleeman2011information, delsole2005predictability, giannakis2012quantifying, kleeman2002measuring, branicki2013non, chen2014information}.

The relative entropy is defined as follows:
\begin{equation}\label{Relative_Entropy}
  \mathcal{P}(p(t),p_{eq}) = \int p(t)\log\left(\frac{p(t)}{p_{eq}}\right),
\end{equation}
which is also known as  Kullback-Leibler divergence or information divergence \cite{kullback1951information, kullback1987letter}.
One practical setup in many applications arises when both the distributions involve only the mean and covariance so that
$p(t)\sim\mathcal{N}(\bar{\mathbf{u}}, \mathbf{R})$ and $p_{eq}\sim\mathcal{N}(\bar{\mathbf{u}}_{eq}, \mathbf{R}_{eq})$
are Gaussian distributions. In such a case, $\mathcal{P}(p(t),p_{eq})$ has an explicit formula
\begin{equation}\label{Signal_Dispersion}
  \mathcal{P}(p(t),p_{eq}) = \left[\frac{1}{2}(\bar{\mathbf{u}}-\bar{\mathbf{u}}_{eq})^*\mathbf{R}_{eq}^{-1}(\bar{\mathbf{u}}-\bar{\mathbf{u}}_{eq})\right] + \left[-\frac{1}{2}\log\det(\mathbf{R}\mathbf{R}_{eq}^{-1}) + \frac{1}{2}(\mbox{tr}(\mathbf{R}\mathbf{R}_{eq}^{-1})-|\mathcal{K}|)\right],
\end{equation}
where  `det' and 'tr' are the determinant and the trace of a matrix and $|\mathcal{K}|$ is the dimension of the system. In \eqref{Signal_Dispersion}, the first term in brackets is called `signal', reflecting the information difference in the mean but weighted by the inverse of the equilibrium variance, $\mathbf{R}_{eq}$, whereas the second term in brackets, called `dispersion', involves only the information distance regarding the covariance ratio, $\mathbf{R}\mathbf{R}_{eq}^{-1}$.

Despite the lack of symmetry, the relative entropy \eqref{Relative_Entropy} has two attractive features. First, $\mathcal{P}(p(t),p_{eq}) \geq 0$ with equality if and only if $p(t)=p_{eq}$. Second, $\mathcal{P}(p(t),p_{eq})$ is invariant under general nonlinear changes of variables.
The signal and dispersion terms in \eqref{Signal_Dispersion} are individually invariant under any (linear) change of variables which maps Gaussian distributions to Gaussians.
In the LEMDA framework with linear stochastic forecast models, both the prior and posterior distributions are Gaussian. Thus, the formula in \eqref{Signal_Dispersion} will be used to quantify the uncertainty reduction. Since $\mathcal{P}(p(t),p_{eq})$ is time-dependent, the skill score shown below is its time average.

\section{Lagrangian DA}\label{Sec:LagrangianDA}
Let us start understanding the Lagrangian component in the LEMDA framework. The results in this Section are purely based on the Lagrangian DA.
\subsection{Performance of the Lagrangian DA}
The underlying flow model is given by \eqref{Ocean_Velocity}--\eqref{OU_process} with double periodic boundary conditions. The flow field is assumed incompressible, and no mean background flow is included. The incompressibility can be imposed on the eigenvectors in \eqref{Ocean_Velocity} \cite{chen2014information}. The parameters adopted in the following experiments are:
\begin{equation}\label{Parameters_LaDA_model1}
  d_\bfs{k} = 1,\qquad \omega_\bfs{k}=0,\qquad {f}_\bfs{k}= \bfs{0} \qquad \mbox{and}\qquad \sigma_\bfs{k} = 0.05,
\end{equation}
for all $\bfs{k}$ such that the flow field has an equipartition of the energy that generates turbulent features. The range of the spectral modes is $\bfs{k}\in[-4,4]^2$. Excluding the background mode $\bfs{k}=(0,0)$, there are in total $80$ modes. Translating the non-dimensional flow field to the dimensional form, the choice of the parameters corresponds to a flow field in the square domain with size being $1000$km$\times1000$km, e-folding time being about 20 days, and the order of velocity being $0.05$m/s (which corresponds to an amplitude of $0.5$ in the non-dimensional form as shown in all the figures). This mimics the typical ocean flow field \cite{taylor2023submesoscale}.

The parameters in the tracer dynamics \eqref{Lagrangian_Tracer_Model} are taken as
\begin{equation}\label{Parameters_LaDA_model2}
\sigma_x = 0.001 \qquad\mbox{and}\qquad\bfs{f}_l = \bfs{0}.
\end{equation}
In such a setup, tracers do not suffer from collision forces. Section \ref{Sec:EulerianDA} will include a comparison of the Eulerian and Lagrangian DA in the presence of collision forces. The noise level here represents the observational error in the satellite image. Converting to the dimensional values, this noise induces an error of about $160$m per day, which is of the order of one-pixel distance (typical $250$m) \cite{manucharyan2022spinning} and is a reasonable choice as the measurement error. The drag coefficient $\beta=1$ is within the range for typical ice floes in the Arctic region \cite{damsgaard2018application}, and the DA skill different values of $\beta$ will be tested in the following numerical simulations. The initial distribution of tracers is uniform. The numerical integration time step is $\Delta{t}=0.0001$. The DA is computed for $t\in[0,10]$, and the skill scores are calculated based on the solution within the time interval $t\in[1,10]$ to eliminate the effect from the burn-in period.

\begin{figure}[ht]
    \includegraphics[width=1.0\textwidth]{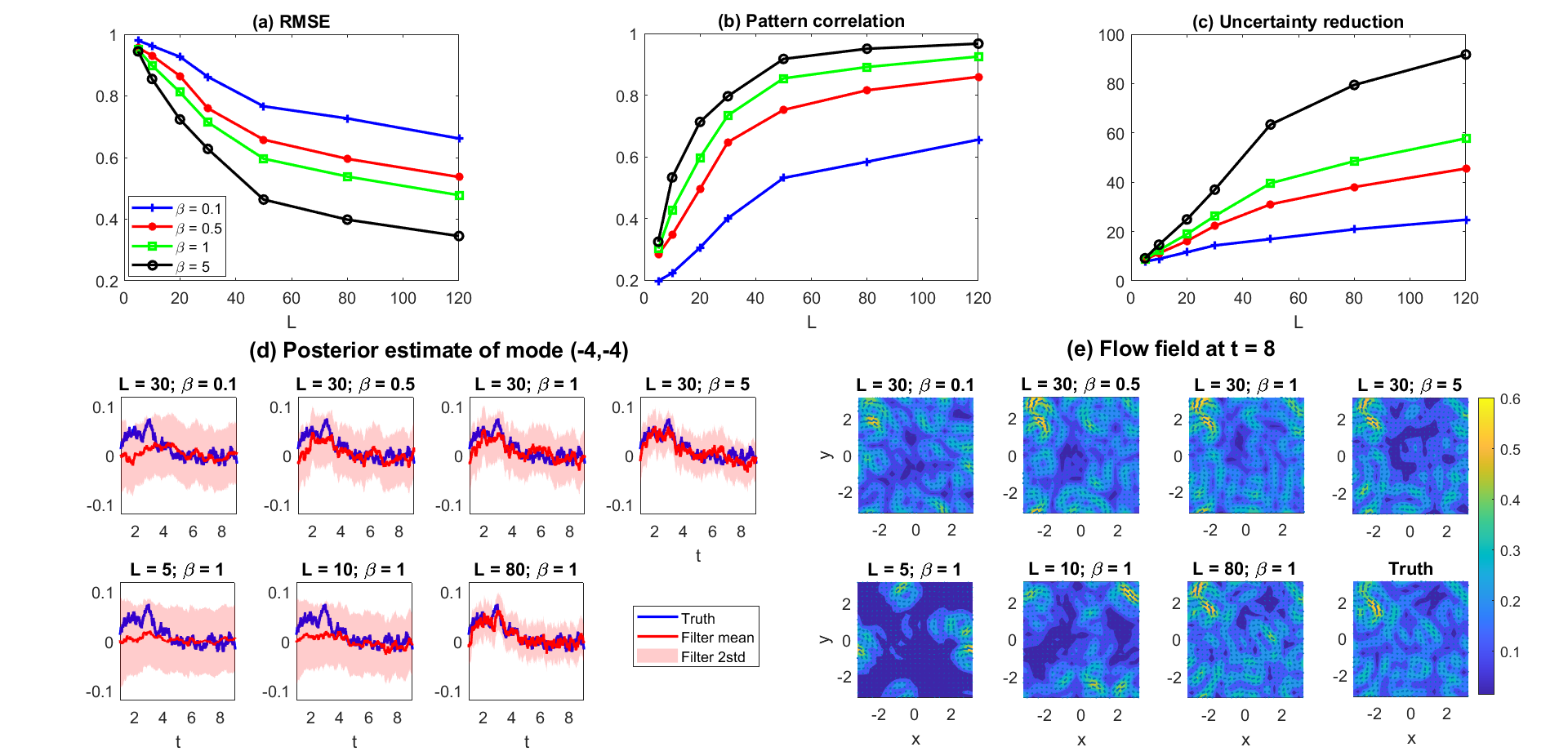}
    \caption{Lagrangian DA with parameters in \eqref{Parameters_LaDA_model1} and \eqref{Parameters_LaDA_model2}. Panels (a)--(c) show the RMSE \eqref{Skill_Scores_RMSE}, pattern correlation \eqref{Skill_Scores_Corr} and uncertainty reduction \eqref{Signal_Dispersion} as a function of the tracer numbers $L$ (x-axis) and drag coefficient $\beta$ (different curves). Panel (d) shows the real part of the posterior estimate of the mode $(-4,-4)$ with different $L$ and $\beta$. Panel (e) shows the flow field at $t=8$ with different $L$ and $\beta$. The last subpanel includes the truth. Note that in this and the subsequent figures, all flow fields are shown in the non-dimensional units. The velocity with an amplitude of $0.5$ corresponds to $0.05$m/s. }
    \label{Figs:LaDA_Scores}
\end{figure}

Figure \ref{Figs:LaDA_Scores} presents the Lagrangian DA results. Panels (a)--(c) show the RMSE \eqref{Skill_Scores_RMSE}, pattern correlation \eqref{Skill_Scores_Corr} and uncertainty reduction \eqref{Signal_Dispersion} as a function of the tracer numbers $L$ (x-axis) and drag coefficient $\beta$ (different curves). There are two main conclusions. First, the error in the posterior mean decreases when the number of observations $L$ increases. Similarly, when $L$ becomes large, the uncertainty reduction becomes more significant, which means the posterior uncertainty shrinks, and the posterior mean estimate is more confident. Second, the DA skill depends on the drag coefficient $\beta$. A smaller drag coefficient leads to a decline in the skill in the state estimation of the flow field. This is not surprising since a smaller $\beta$ implies a smaller signal-to-noise ratio in \eqref{Lagrangian_Tracer_Model_v}. In other words, the observability suffers from the decrease of $\beta$ when DA applies to recover the flow field. Panels (d) and (e) show the posterior estimate of the mode $(-4,-4)$ and a snapshot of the flow field in physical space with different $L$ and $\beta$. The results in these two panels support the conclusions drawn from the skill scores in Panels (a)--(c).

\subsection{Reduced-order Lagrangian DA}\label{Subsec:Reduced_LDA}
Despite the analytic formulae \eqref{eq:filter}, potential computational challenges in the Lagrangian DA may appear when the dimension of the flow field becomes large. Recall that the posterior covariance matrix scales as a quadratic function of the dimension of the flow, which is the total number of spectral modes in this paper. Therefore, the major computational burden comes from solving the matrix Riccati equation \eqref{eq:filter_R} for the time evolution of the posterior covariance. Effective approximations in accelerating the estimation of the posterior covariance will significantly facilitate the entire DA procedure.

If the tracers are nearly uniformly distributed in the domain, then an approximate scheme for reducing the computational cost in solving the posterior covariance matrix can be derived. Note that if the flow field is nearly incompressible and the drag coefficient $\beta$ is not too small, then the tracers are unlikely to be highly concentrated in a particular area \cite{chen2014information}. Therefore, such a requirement of the tracer distribution is reasonable in many ocean and fluid applications. One essential basis for reaching this approximate scheme is the following. Although different spectral modes are fully coupled in the observational processes via \eqref{Ocean_Velocity}, the linear stochastic models of these modes are by design independent in the forecast model \eqref{CGNS_U}. This is one of the crucial features of DA in spectral space, where the strong spatial dependence is primarily described by the spatial bases. This conclusion generally holds even when the governing equations of the spectral modes are coupled due to being decomposed from a complicated partial differential equation \cite{majda2012filtering}. Therefore, one natural strategy for developing approximate DA is to set the part associated with the flow modes in the posterior covariance \eqref{eq:filter_R} to be a diagonal matrix. This significantly reduces the computational cost, since only the diagonal entries need to be updated. Furthermore, these diagonal components are near constants in Lagrangian DA and can be predetermined.

Denote by $R_{\bfs{k}}$ the $i$-th diagonal component of the covariance in \eqref{eq:filter_R} associated with the $i$-th spectral mode of the flow field. By applying the mean-field theory \cite{chaikin1995principles} with the assumption that the tracers are nearly uniformly distributed in the field, its governing equation can be written as
\begin{equation}\label{Rk}
\frac{\d R_{\bfs{k}}}{\d t}= -2d_\bfs{k}R_\bfs{k} +\sigma_\bfs{k}^2 - \sigma_x^{-2}\xi^2 L R_{k}^2,
\end{equation}
where $\xi$ is a parameterization that can be derived.
The positive root of $R_{\bfs{k}}$ in \eqref{Rk} corresponds to the constant approximate solution of the covariance, which is given by
\begin{equation}\label{root_Rk}
  R_{\bfs{k}} = \frac{-d_k + \sqrt{d_k^2+\sigma_x^{-2}\xi^2L\sigma_k^2}}{\sigma_x^{-2}\xi^2L}.
\end{equation}
See the Supporting Information for details. The implication of the result in \eqref{root_Rk} is that the covariance of the flow field no longer needs to be solved via solving the matrix Riccati equation, which significantly reduces the computational cost.

\begin{figure}[ht]
    \includegraphics[width=1.0\textwidth]{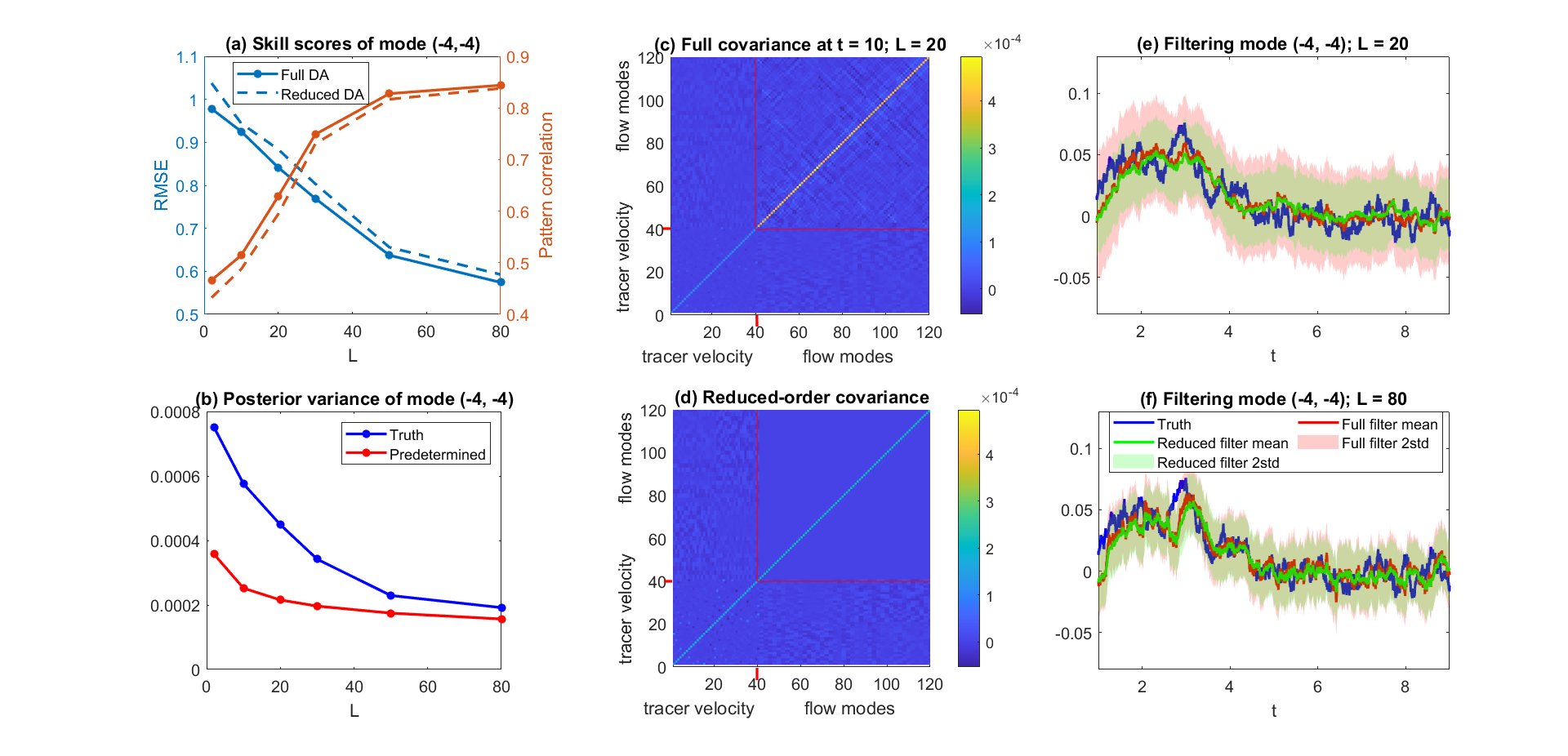}
    \caption{Lagrangian DA and its reduced-order form, where $\beta=1$ is adopted. Panel (a) compares the skill scores of the full and reduced-order DA solutions in terms of the normalized RMSE and the pattern correlation of mode $(-4,-4)$. Panel (b) shows the diagonal entry of the posterior variance of mode $(-4,-4)$ resulting from the full DA scheme averaged over the time interval $[1,10]$ with that given by the predetermined value in \eqref{root_Rk_SI}. Panel (c) shows the posterior covariance from the filtering solution at the last time instant $t=10$ with $L=20$ tracers. Panel (d) shows the posterior covariance when the reduced-order DA scheme is applied. Panels (e)--(f) demonstrate the posterior mean time series and the posterior uncertainty for mode $(-4,-4)$. The posterior uncertainty is given by the two standard deviations from the mean value. }
    \label{Figs:Reduced_LDA}
\end{figure}

Figure \ref{Figs:Reduced_LDA} shows the results of the Lagrangian DA using the exact analytic solution in Section \ref{Subsec:Closed_Formulae} and its reduced-order form in \eqref{root_Rk}, where the latter exploits a pre-determined constant diagonal matrix for the posterior covariance of the part associated with the underlying flow field. Panel (a) compares the skill scores of the full and reduced-order DA solutions. The two DA schemes behave similarly, especially when the number of tracers increases. This is consistent with the condition of the mean-field theory appearing in the derivation shown in the Supporting Information. The result here demonstrates the skilful behaviour of the cheap reduced-order DA scheme. Panel (b) shows the diagonal entry of the posterior variance of mode $(-4,-4)$ resulting from the full DA scheme averaged over the time interval $[1,10]$ with that given by the predetermined value in \eqref{root_Rk}. Note that since the equipartition of the energy is adopted here, other modes behave similarly to this one. The predetermined diagonal entry becomes more consistent with the actual value when $L$ increases. This also provides evidence that the two DA schemes lead to comparable results. Panel (c) shows the posterior covariance from the filtering solution at the last instant $t=10$ with $L=20$ tracers. In such a case, the covariance associated with the flow field (the top right part with the boundaries given by the thin red lines) is already diagonally dominant. Therefore, once the reduced-order DA is applied in Panel (d), the results of the other part of the covariance matrix remain similar. Finally, Panels (e)--(f) demonstrate the posterior mean time series and the posterior uncertainty for mode $(-4,-4)$. Even with $L=20$, the solutions are close to each other. When $L$ increases to $L=80$, the two solutions almost overlap and approach the truth.

\section{Eulerian DA}\label{Sec:EulerianDA}

This section includes the Eulerian DA results with two different scenarios. In the first case, the collision force is omitted, and the goal is to understand the skill of the Eulerian DA using the continuum equations derived in Section \ref{Subsec:Eulerian_Model} as the forecast model. In the second scenario, ice floes are utilized as tracers. Therefore, collisions due to the floe-floe contact forces play a vital role in driving the motions of the floes, which significantly affects the direct Lagrangian DA. The skill of the Eulerian DA in recovering large-scale features of the flow field will be illustrated.

\subsection{Eulerian DA with no particle collisions}\label{sec:eudanum}
Let us start with the same case as in Section \ref{Sec:LagrangianDA}, where the flow and the particle models are equipped with the parameters in \eqref{Parameters_LaDA_model1}--\eqref{Parameters_LaDA_model2} with $80$ spectral modes. In addition, the drag force is $\beta=1$.

The performance of the Eulerian DA with a rather coarse grid of size $9\times 9$ for recovering the velocity of the underlying flow field is studied. A total of $L=2000$ particles in the flow field are utilized to compute the averaged momentums of particles in each grid cell as observational quantities. On average, each grid cell has about $25$ particles. As will be seen later, this number is sufficient to reach a skilful DA result. It is worth highlighting that the computational cost of calculating these averaged momentums is nearly negligible compared with using a large number of particles as the direct observations for Lagrangian DA. Therefore, there should not be any concerns regarding using such a large number of particles (if available) in the Eulerian DA from the computational cost aspect. The governing equations of the momentums are given by \eqref{eq:mv}. The particle velocity is computed using a finite difference scheme based on the displacement in the two consecutive time instants that differed by $\Delta t$. The velocity of each particle suffers from random noise, as seen in \eqref{Lagrangian_Tracer_Model_x}. Nevertheless, the statistical average will eliminate the effect of the random noise to a large extent. In the Eulerian DA, the observed variables are the momentums in these grid cells. They are denoted by $\bfs{Z}$ in \eqref{CGNS_X}. For a grid of size $9\times 9$, the noise in the observational processes $\bfs{Z}$ is given by $\sigma_z = 0.02$, where $\sigma_z$ is the diagonal component of the identity matrix $\bfs{\Sigma}_\bfs{Z}$ in \eqref{CGNS_X}. The value $\sigma_z$ is computed based on the procedure described in subsection \ref{sec:eudanoise} of the Supporting Information. As in the Lagrangian DA case, the tracers are uniformly distributed over the domain $[-\pi, \pi]^2$ at the initial time instant.

\begin{figure}[ht]
    \includegraphics[width=1.0\textwidth]{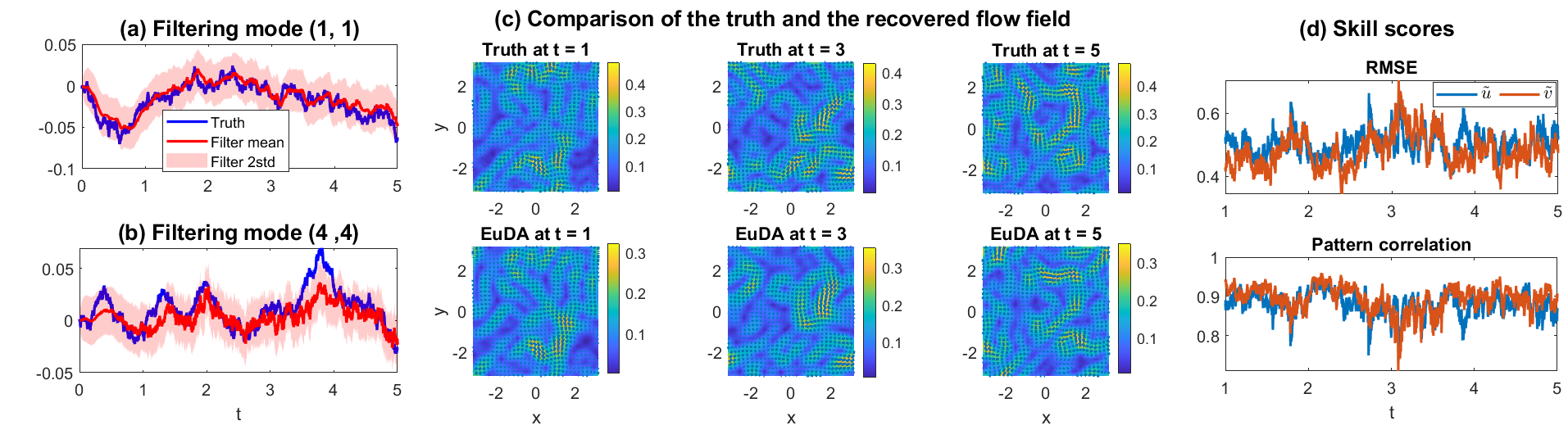}
    \caption{Eulerian DA of the flow field using without particle collisions. The observations of the momentums are computed based on $L = 2000$ observed particle and the Eulerian observations are taken at a $9\times 9$ mesh grid. Panels (a)--(b) compare the filtering results with the truth for two Fourier modes $(1,1)$ and $(4,4)$. Panel (c) shows the comparison of the truth and the recovered velocity field in physical space at $t=1,3$ and $5$. Panel (d) shows the skill scores of the velocity field in physical space.
    }
    \label{fig:EuDA}
\end{figure}

Figure \ref{fig:EuDA} presents the numerical results. Panels (a)--(b) show the filtering results in recovering two Fourier modes. The posterior mean captures the patterns of the truth quite well with relatively moderate uncertainty. Panel (c) compares the truth and the recovered velocity field in physical space at $t=1,3$ and $5$. These results further provide an intuitive understanding that the Eulerian DA is skilful in capturing the features in the true flow field. Panel (d) shows the skill scores of the velocity field in physical space, which are quantitative indicators that support the findings in the previous panels. Notably, the EuDA, even with such a coarse grid, can perform well with a pattern correlation over $0.85$.

\begin{figure}[ht]
\includegraphics[width=1.0\textwidth]{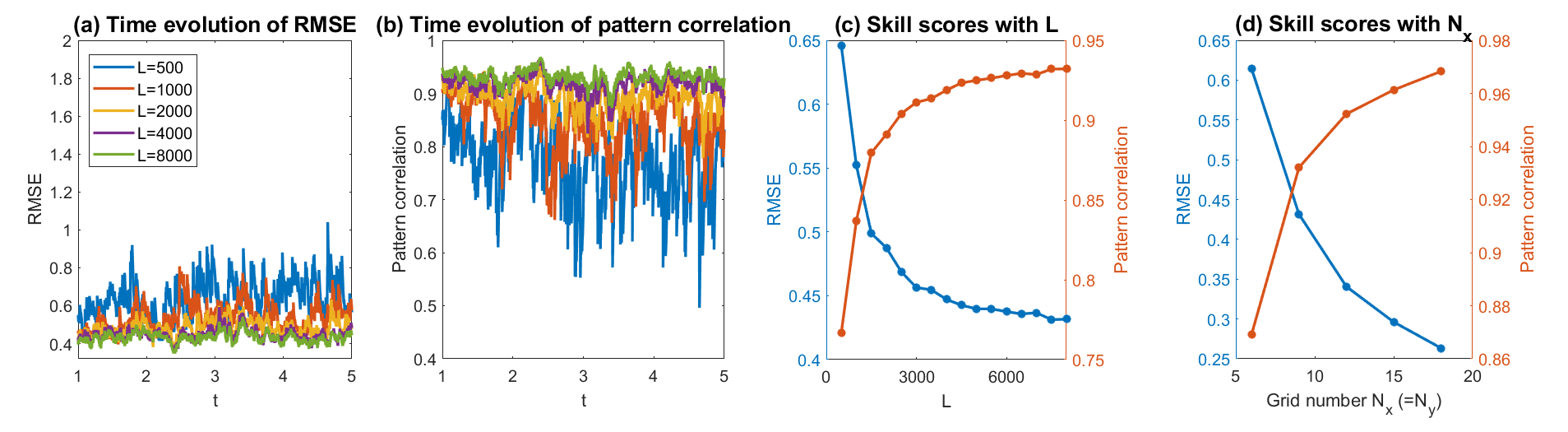}
    \caption{Skill scores of Eulerian DA with different number of particles $L$ and grid size $N_x=N_y$. Panels (a)--(b) show the RMSE and pattern correlation as a function of time using different $L$, where the mesh size $N_x=N_y=9$ is fixed. Panel (c) shows the skill scores average over time as a function of $L$. Panel (d) shows the skill scores as a function of $N_x=N_y$, where $L=8000$ is used.
    }
    \label{fig:EuDAerr}
\end{figure}

Figure \ref{fig:EuDAerr} provides an understanding of the impacts of the number of particles $L$ and the grid sizes on the performance of the Eulerian DA. For simplicity, equal numbers of the grid points in the $x$ and the $y$ directions are taken, namely $N_x=N_y$. Panels (a)--(b) show the time evolutions of the RMSE and the pattern correlation using different $L$, where the mesh size $N_x=N_y=9$ is fixed. The time-averaged skill scores are shown in Panel (c). As expected, the skill scores improve as the number of particles used to calculate the observational quantities (i.e., the momentums) at fixed grid cells increases. The least accurate case in these panels corresponds to $L =500$. Since a $9\times9$ mesh grid is adopted, each cell has, on average, only six particles. This reduces the accuracy of DA. In particular, the cells contain no particles from time to time. In such a case, the observed values are interpolated using the computed momentum values in the nearby grids. The details are contained in Subsection \ref{Subsec:Interpolation} in the Supporting Information. Nevertheless, even using such a small number of particles to calculate the momentums, the pattern correlation is above $0.7$, and the RMSE is, on average, below $0.8$, which means the DA remains relatively skilful. This indicates the robustness of the method. Next, Panel (d) shows the skill scores as a function of $N_x=N_y$, where $L=8000$ is used. The skill scores improve as the grid size is refined. Notably, when the grid size is as coarse as $6\times 6$, the pattern correlation remains above $0.85$. This provides a reasonable justification for using the Eulerian DA to recover the large-scale features of the underlying flow field.

\subsection{Recovery of large-scale flow feature using Eulerian DA in the presence of particle collisions}\label{Subsec:Collisions}

Consider a case where the force in the tracer equation \eqref{Lagrangian_Tracer_Model} is nonzero, i.e., $\bfs{f}_l \ne 0$. This appears, for example, when ice floes are adopted as tracers in recovering the underlying ocean field in the marginal ice zone \cite{mu2018arctic, lopez2023level}. To this end, the particles are also named ice floes in this subsection. For simplicity, the floes are given by disks of uniform size and thickness. Therefore, the force $\bfs{f}_l$ characterizes the particle-particle collision force. Hooke's law of elasticity is naturally used to model this collision force \cite{manucharyan2022subzero, damsgaard2018application}. Decompose the total force into two parts:
\begin{equation}
    \bfs{f}_l = \sum_{j\in J}(\bfs{f}_{\bfs{n}}^{lj} + \bfs{f}_{\bfs{t}}^{lj}),
\end{equation}
where $\bfs{f}_\bfs{n}^{lj}$ and $\bfs{f}_\bfs{t}^{lj}$ are the normal and tangential components, respectively, from the $j$-th floe in the set $J$. The set $J$ contains all the floes that contact the $l$-th floe at a certain time instant. A simplification is made. Assume that the floes only have translational momentums, but the angular momentum is omitted. This is sufficient for illustrating the key differences in the Eulerian and Lagrangian DAs. In such a case, the contact forces are given by
\begin{equation}\label{Youngsmodulus}
\begin{aligned}
     \bfs{f}_{\bfs{n}}^{lj} & = c^{lj}E^{lj}\delta^{lj}\bfs{n}^{lj},\qquad
     \delta_n^{lj} \equiv d^{lj} - 2r,
    \quad\mbox{with}\quad d^{lj} = |\bfs{x}^l-\bfs{x}^j|,\\
     \bfs{f}_{\bfs{t}}^{lj} & = c^{lj}G^{lj}v_{\bfs{t}}^{lj}\bfs{t}^{lj}, \qquad v_{\bfs{t}}^{lj} = (\bfs{v}^j- \bfs{v}^l )\cdot\bfs{t}^{lj},
\end{aligned}
\end{equation}
where for simplicity all the floes are assumed to have the same radius $r= 0.01\pi$ (corresponding to the same area $S=\pi r^2 = \pi^3\times 10^{-4}$). In \eqref{Youngsmodulus}, $E^{lj}$ is the Young's bulk modulus, $G^{lj}$ is the Young's shear modulus, and $c^{lj}$ is the chord length in the transverse direction of the cross-sectional area \cite{cundall1979discrete}. The vectors $\bfs{n}$ and $\bfs{t}$ are the unit vectors along the normal and the tangential directions. The Coulomb friction law is further applied to limit the tangential contact force relative to the magnitude of the normal contact force. The restriction $|\bfs{f}_\bfs{t}^{lj}| \leq \mu^{lj}|\bfs{f}_\bfs{n}^{lj}|$ is applied, where $\mu^{lj}$ is the coefficient of friction that characterizes the condition of the surfaces of the two floes in contact. In the following simulations, $\mu^{lj} = 0.2$ is adopted. By choosing reasonable values of Young's moduli, the amplitude of the collision forces is comparable to that of the drag forces.

The same setup as that in subsection \ref{sec:eudanum} is utilized here except that the collisions are further added in this experiment. The test regime here contains 2000 floes, which corresponds to a concentration of about $0.16$. This concentration allows the floes to trigger relatively frequent contact with each other.

\subsubsection{Issues in applying the Lagrangian DA}\label{Subsec:LaDA_collisions}
\begin{figure}[ht]
    \includegraphics[width=1.0\textwidth]{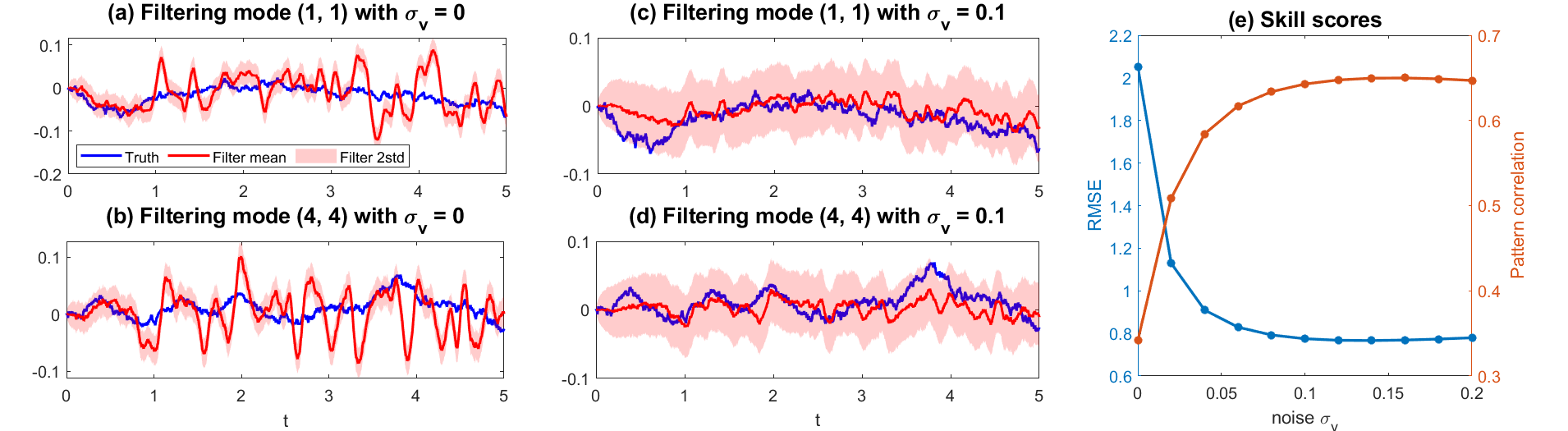}
    \caption{Lagrangian DA by randomly selecting 40 floes from in total 2000 floes in the domain. Here, the full Lagrangian DA rather than the reduced-order version is utilized. Panels (a)--(b) show the filtering results of mode $(1,1)$ and $(4,4)$ using the original forecast model by ignoring the contact force and without additional noise, that is, $\sigma_v = 0$ in \eqref{Lagrangian_Tracer_Model_v_modified}. Panels (c)--(d) show the results but inflate the noise in the governing equation of the floe velocity by taking $\sigma_v = 0.1$. Panel (e) shows} the skill scores as a function of the level of noise inflation.
    \label{fig:cfLaDAnp40}
\end{figure}

It is worth highlighting that collisions occur essentially instantaneously. Therefore, a tiny numerical integration time step is needed to resolve the collisions in the Lagrangian DA, affecting computational efficiency. Such an issue can be avoided when the Eulerian DA is adopted, which will be presented in Section \ref{Subsec:EulerianDA_collsions}. In addition, perfectly resolving collisions is inaccessible in practice. A small bias in describing the contact force due to the approximation in geometric properties or the collision law may lead to a significant error in the Lagrangian DA. Even if the contact force is perfectly characterized, when all the 2000 floes are utilized as the observed trajectories, the Lagrangian DA becomes quite expensive as the dimension of the state variable shoots up. To remain an efficient DA, only 40 floes are randomly selected as observations. As a majority of the floes are not included as observational variables, the contact force is hard to accurately characterize in the forecast model. Thus, it is ignored in the forecast model of the floe velocity, which simply becomes
\begin{equation}\label{Lagrangian_Tracer_Model_v_no_force}
  \frac{\d \bfs{v}_l}{\d t} = \beta(\bfs{u}-\bfs{v}_l).
\end{equation}
Such a setup serves as the benchmark. Panels (a)--(b) of Figure \ref{fig:cfLaDAnp40} show the filtering results of mode $(1,1)$ and $(4,4)$ using the original forecast model in \eqref{Lagrangian_Tracer_Model_v_no_force} by ignoring the contact force. Vast oscillations are seen in the posterior mean time series. These large biases happen when collisions occur at these time instants while the contact force is ignored in the forecast model. The RMSE is twice as much as the standard deviation of the model equilibrium and the pattern correlation is only 0.35, both of which clearly indicate that the Lagrangian DA loses its skill in recovering the underlying flow field.

One possible way to improve the results of the Lagrangian DA is to apply the noise inflation (also known as covariance inflation) strategy, which is widely utilized in practice \cite{hamill2005accounting, silva2021new}. To this end, the forecast model of the floe velocity \eqref{Lagrangian_Tracer_Model_v} is modified by
\begin{equation}\label{Lagrangian_Tracer_Model_v_modified}
  \frac{\d \bfs{v}_l}{\d t} = \beta(\bfs{u}-\bfs{v}_l) +\sigma_v\dot{\bfs{W}}_l.
\end{equation}
The additional noise effectively characterizes the uncertainty due to the collisions. It can be regarded as modelling the collision effects in a statistical way in the forecast model. Panels (c)--(d) show the DA results by inflating the noise in the forecast model of the floe velocity by taking $\sigma_v = 0.1$. The posterior estimate is significantly improved. Panels (e)--(f) show the skill scores as a function of the level of noise inflation, which provides a quantitative indication of the improvement of the DA skill. With noise inflation from $\sigma_v = 0.0$ to $\sigma_v = 0.1$, the pattern correlation increases from Corr=$0.35$ to Corr=$0.65$. Yet, the improved pattern correlation is still below the one associated with the case using a perfect forecast model (Corr=$0.80$) shown in Figure \ref{Figs:LaDA_Scores}. These results confirm that noise inflation can help improve the results to a large extent, but a barrier still exists by utilizing such a statistical characterization of the contact force.

\subsubsection{Improving the results using Eulerian DA}\label{Subsec:EulerianDA_collsions}

\begin{figure}[ht]
    \includegraphics[width=1.0\textwidth]{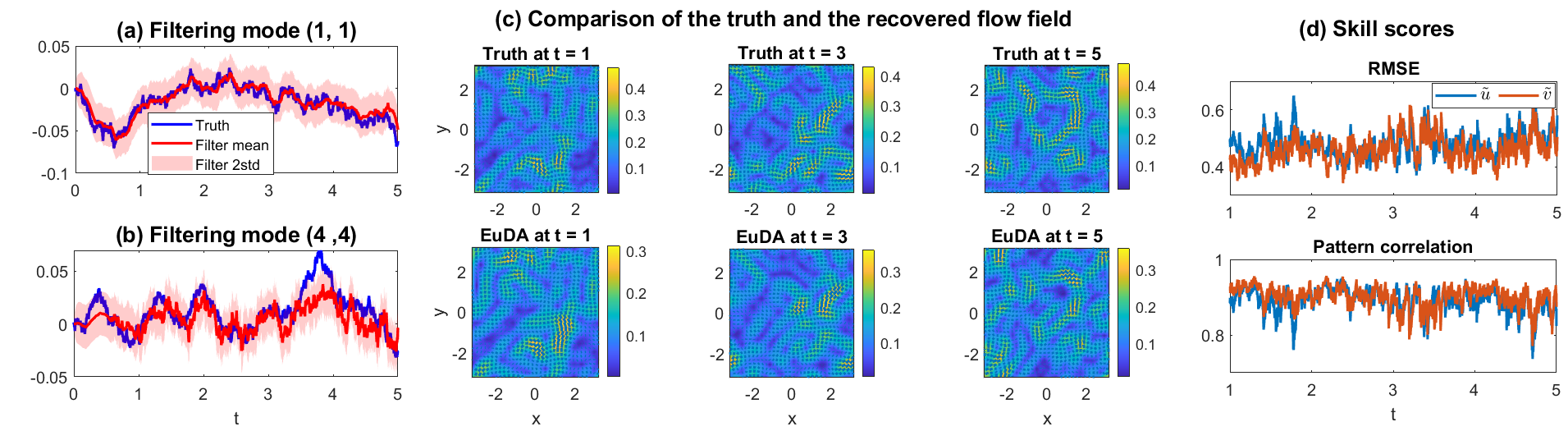}
    \caption{Eulerian DA of the flow field using particle collisions. The observations of the momentums are computed based on $L = 2000$ observed particle and the Eulerian observations are taken at a $9\times 9$ mesh grid. Panels (a)--(b) compare the filtering results with the truth for two Fourier modes $(1,1)$ and $(4,4)$. Panel (c) shows the comparison of the truth and the recovered velocity field in physical space at $t=1,3$ and $5$. Panel (d) shows the skill scores of the velocity field in physical space.
    }
    \label{fig:cfEuDA}
\end{figure}

Figure \ref{fig:cfEuDA} shows the results using the Eulerian DA based on the $L=2000$ floes and a mesh with $9\times9$ grid cells. The posterior estimates from the Eulerian DA are much more accurate than those from the Lagrangian DA in Section \ref{Subsec:LaDA_collisions}. The underlying reason is that the effect of collisions is averaged out when the statistical moments, i.e., the momentums, are computed. Notably, the Eulerian DA shown in Figure \ref{fig:cfEuDA} leads to comparable results as the case without particle collision shown in Figure \ref{fig:EuDAerr}.

\begin{figure}[ht]
    \includegraphics[width=1.0\textwidth]{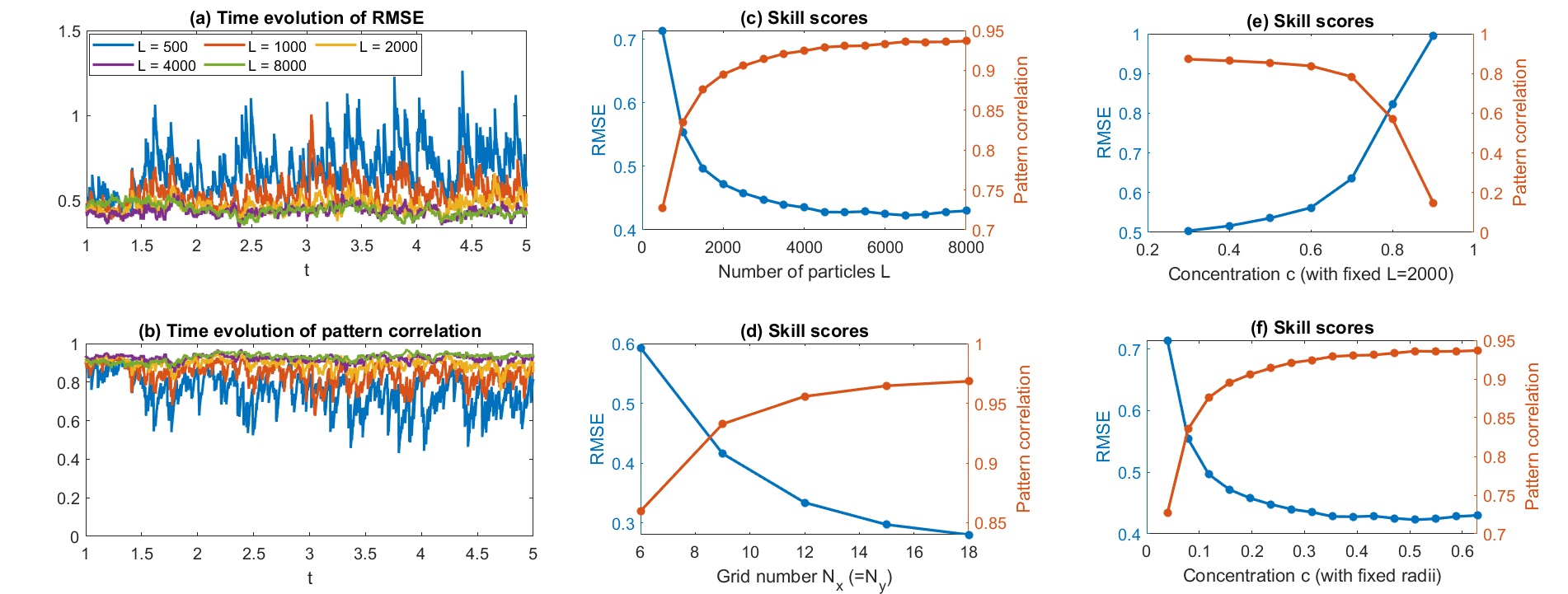}
    \caption{Skill scores of EuDA with particle collisions. Panels (a)--(b) show the skill scores as a function of time with different $L$ where the particle radii are fixed as $0.01\pi$. With the same setup, the time-averaged skill scores are shown in Panel (c). Panel (d) shows the skill scores as a function of the mesh grids $N_x=N_y$, where the total number of floe is $L=8000$, and the concentration is fixed at $c=0.63$. Panel (e) shows the skill scores as a function of concentration, where the total number of floes $L=2000$ is fixed, but the radius varies to reach different concentration values. Note that the maximum concentration is above $\pi/4$ as small overlaps between different floes are allowed when the elastic collision formula is applied \cite{cundall1979discrete}. Panel (f) shows the skill scores as a function of the concentration, where the radius is fixed and the number of floes $L$ varies.
    }
    \label{fig:cfEuDAerr}
\end{figure}

Figure \ref{fig:cfEuDAerr} studies the robustness of the Eulerian DA in the presence of particle collisions. Panels (a)--(b) show the skill scores as a function of time with different $L$ where the particle radii are fixed as $0.01\pi$.
 The time-averaged skill scores are shown in Panel (c). Similar to the case without collision, with an increased number of floes, the Eulerian DA has smaller statistical errors and leads to more accurate state estimations. Panel (d) shows the skill scores as a function of the mesh grids $N_x=N_y$, where the total number of floe is $L=8000$, and the concentration is fixed at $c=0.63$. The results indicate that a more refined grid helps eliminate the discretization error.  Panel (e) shows the skill scores as a function of concentration, where the total number of floes $L=2000$ is fixed, but the radius varies to reach different concentration values. When the concentration increases and exceeds a certain threshold of about $0.65$, the skill scores decline quickly. The reason is that when the floes are more packed, they become hard to move around. In other words, the contact force becomes more dominant than the drag force. The observability of the ocean is thus significantly affected. In the limit that the domain is fully packed, all the floes are static, which becomes useless in recovering the underlying flow field. Finally, Panel (f) shows the skill scores as a function of the concentration, where the radius is fixed and the number of floes $L$ varies. It is seen that before arriving at the threshold of declining performance, the skill scores remain robust.

\section{LEMDA for Recovering Multiscale Turbulent Flows}\label{Sec:lemda}
In this section, the two components will be combined to facilitate the DA of a multiscale turbulent flow field. The general description below assumes the particles have no collisions. In practice, non-interacting floes can naturally be selected to provide observed trajectories for implementing Lagrangian DA.


\subsection{The procedure of applying LEMDA}
LEMDA consists of two steps. The first step is to apply the Eulerian DA to recover the large-scale modes in a relatively coarse grid. The second step is to apply the Lagrangian DA with a small number of observed particles in each coarse grid cell to recover the small-scale modes, where the large-scale modes are treated as known modes obtained from the first step.

The following crucial features of LEMDA are worth highlighting. First, as the Eulerian DA is only applied to a relatively coarse grid, there are sufficient numbers of particles to compute the observed statistical quantities, which allows an accurate DA result. Meanwhile, since the grid is coarse, computational efficiency is guaranteed. Second, fundamentally different from the expensive direct implementation of the Lagrangian DA to the entire domain, the Lagrangian DA in the LEMDA framework only applies to each grid cell. Notably, the Lagrangian DAs at different grid cells are implemented independently. In other words, parallel computing can naturally be applied for carrying out these Lagrangian DAs, where no message passes across different processors. This parallel DA procedure using Lagrangian observations is highly efficient, especially since only a few observations are used to recover the small-scale flow structures in each grid cell.

\subsection{Implementation details}
In the first step, when the Eulerian DA is applied to recover the large-scale features of the flow field, the small-scale modes are treated as unresolved in DA. As the spectral method is used to characterize the flow field, the contribution of the strong signals of the small-scale modes averaged over a coarse grid cell is insignificant overall. This allows the Eulerian DA to recover the large-scale features skillfully.

In the second step, when Lagrangian DA is applied in each grid cell, the original system is written into the following form to facilitate the DA:

\begin{subequations}\label{LEMDA:la_model}
\begin{align}
  \frac{\d { \bfs{x}_l}}{\d t} &= \bfs{v}_l + \sigma_x\dot{\bfs{W}}_l, \\
  \frac{\d \bfs{v}_l}{\d t} &= \beta(\bfs{u}_f + \bfs{u}_c -\bfs{v}_l), \qquad\bfs{u}_f(\bfs{x},t) = \sum_{\bfs{k}\in\mathcal{K}, |\bfs{k}| \ge \eta }\hat{u}_{\bfs{k}}(t)e^{i\bfs{k}\bfs{x}}\bfs{r}_{\bfs{k}},\\
  \frac{\d\hat{u}_{\bfs{k}}}{\d t} &= - d_{\bfs{k}}  \hat{u}_{\bfs{k}} + f_{\bfs{k}}(t) + \sigma_{\bfs{k}}\dot{W}_{\bfs{k}},
\end{align}
\end{subequations}
where
$$
\bfs{u}_c = (\bfs{x},t) = \sum_{\bfs{k}\in\mathcal{K}, |\bfs{k}| < \eta }\hat{u}_{\bfs{k}}(t)e^{i\bfs{k}\bfs{x}}\bfs{r}_{\bfs{k}}
$$
is the coarse-scale flow components recovered by the Eulerian DA. In \eqref{LEMDA:la_model}, $\eta$ is pre-defined that separates the large- and small-scale modes. The Lagrangian DA is applied to each coarse cell to recover the small-scale modes $\bfs{u}_f$ (where the subscript $f$ stands for the fine grids). It is worth clarifying that the recovered spectral modes associated with $\bfs{u}_f(\bfs{x},t)$ in \eqref{LEMDA:la_model} in each grid cell only describe the small-scale features within that cell. To obtain the spectral modes characterizing the small-scale features in the entire domain, the local spectral modes from each Lagrangian DA are first utilized to reconstruct the flow field in physical space in each cell. Then, the recovered flow field associated with the small-scale structures in different cells is collected to reach the small-scale flow field in physical space. A new spectral decomposition of this field is finally applied to obtain the spectral modes describing the small-scale features in the entire domain.

In principle, one Lagrangian tracer can go from one cell to another as time evolves. This will not affect the Lagrangian DA as long as the posterior mean and covariance sizes change adaptively in time according to the number of particle observations in each domain. In the simulation here, a simplified setup is utilized. In each grid cell, ten particles are preselected (typically, they are chosen near the centers of the grid cells). These ten particles will stay in the same cell during the DA test period, which allows the posterior mean and covariance to stay in the grid of fixed size.

\subsection{Numerical experiments}
The flow field considered in this numerical simulation is characterized by spectral modes $\bfs{k}\in[-8,8]^2$, and there are in total $288$ modes (the $(0,0)$ mode is excluded). An equipartition of the energy spectrum is again adopted that induces a strongly turbulent system, making the test here quite challenging. Except that the noise coefficients in characterizing the Fourier modes are reduced to $\sigma_\bfs{k} = 0.01$, all the other parameters remain the same as those in \eqref{Parameters_LaDA_model1}--\eqref{Parameters_LaDA_model2}. The purpose of adopting a reduced noise coefficient level is to guarantee that the order of the velocity in physical space remains around $0.05$m/s, which mimics the realistic situation. There are, in total, $3136$ particles in the entire domain with no collision between different particles.

The course grid utilized for the Eulerian DA has a size of $7 \times 7$, the degree of freedom of which is $49$. Thus, it is natural to use the Eulerian DA to estimate the state of the modes with $|\bfs{k}| < 4$ (in total $44$ modes), which are referred to as large-scale modes. The remaining 244 modes with $|\bfs{k}| \ge 4$ are called small-scale modes, which are recovered using the Lagrangian DA.

\begin{figure}[ht]
    \includegraphics[width=1.0\textwidth]{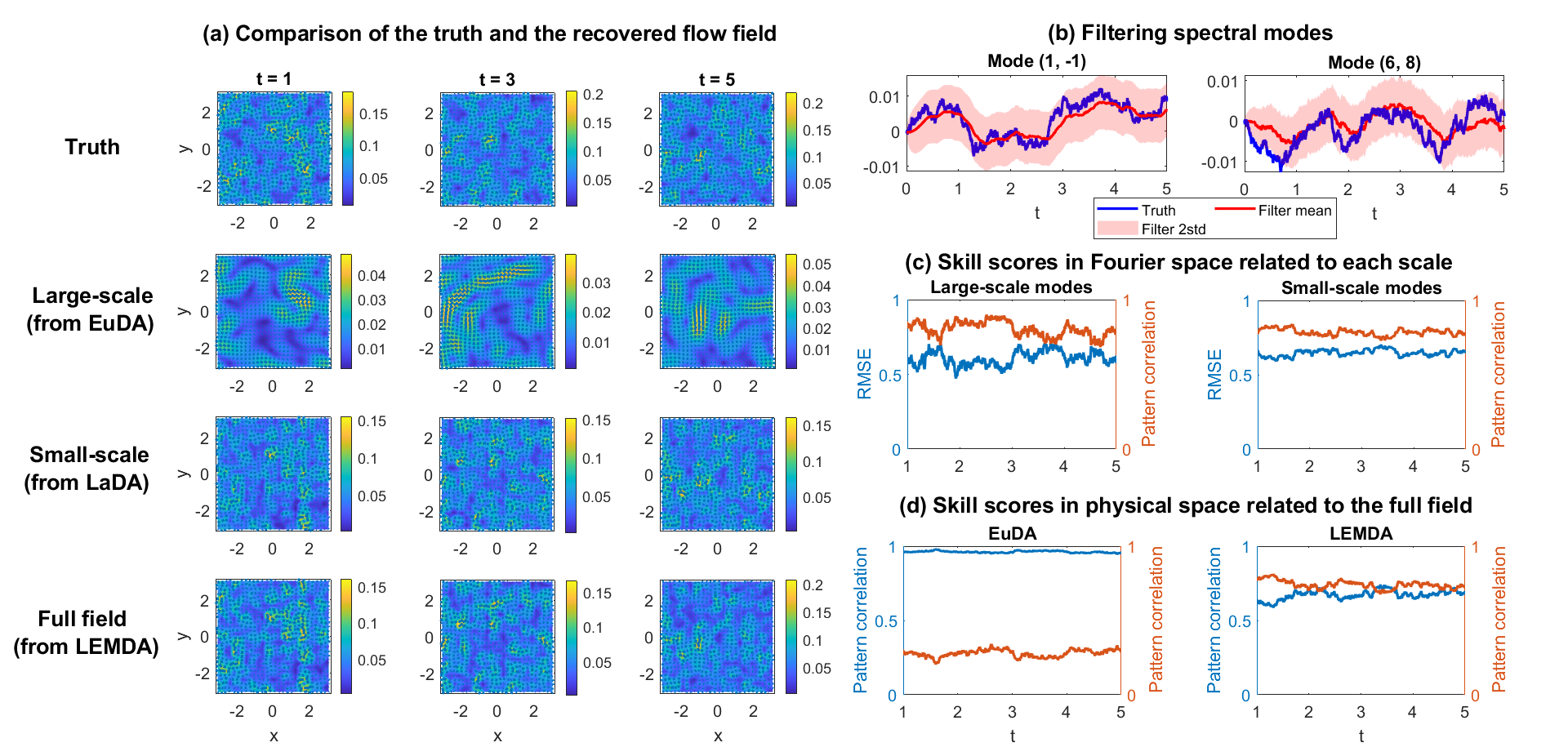}
    \caption{Recovery of multiscale turbulent flow field utilizing LEMDA. The large-scale modes (rough $|\bfs{k}|<4$) are recovered by the Eulerian DA in a coarse grid with size $7\times7$. The small-scale modes (roughly $|\bfs{k}|\ge4$) are recovered by the Lagrangian DA, where 10 particle trajectories in each coarse grid cell are utilized as observations. There are, in total, 3136 particles in the entire domain. Panel (a) compares the truth and the recovered flow fields. Panel (b) shows the filtering results of one large-scale mode $(1,-1)$ and one small-scale mode $(6,8)$. Panel (c) shows the skill score in Fourier space averaged over all the large-scale modes (left) and small-scale modes (right), related to the true large- and small-scale modes, respectively. Panel (d) shows the skill score in physical space associated with the entire flow field for the recovered large-scale features using the Eulerian DA (left) and in the recovered entire flow field using LEMDA (right).
    } \label{fig:lemdakmax8}
\end{figure}

Figure \ref{fig:lemdakmax8} includes the results of LEMDA and the associated Eulerian and Lagrangian DA components in recovering. Panel (a) compares the truth and the recovered flow fields at $t=1, 3$, and $5$. The first row demonstrates the true flow field. As the equipartition of the energy is utilized, the true flow field is highly turbulent. The second row shows the recovered large-scale modes using the Eulerian DA. Note from the colour bar that the amplitude (and thus the energy) of the large-scale modes is only a portion of the total flow field. The filtering solution of one of the large-scale Fourier modes $(1,-1)$ is shown in the left column of Panel (b), which indicates the significant skill of the Eulerian DA. The third row in Panel (a) shows the flow structures associated with the small-scale modes recovered from the parallel Lagrangian DA. The skill of the Lagrangian DA in recovering these refined features can be quantitatively seen in the right column of Panel (b). Summing up the two components of the flow field in the second and third rows in Panel (a) leads to the results in the fourth panel, which is the recovered entire flow field. Comparing the first and the fourth rows, the recovered flow field reproduces the multiscale features of the truth. Next, Panel (c) provides a qualitative understanding of the skill scores in recovering the large- and small-scale features of the flows. Both components are recovered accurately, with pattern correlations above $0.75$. The large-scale features are recovered slightly more accurately than the small-scale components in this case, although the skill of the latter can be enhanced when the number of the observed Lagrangian trajectories increases. The right column of Panel (d) studies the performance of the entire flow field, which is the superposition of the large- and small-scale structures. Again, the pattern correlation is above $0.75$, and the RMSE is below $0.7$, consistent with the intuitions obtained from the first and the fourth rows in Panel (a). Note that the small-scale modes account for a large portion of the total signal. If they are omitted, then the skill score drops significantly, as seen in the left column of Panel (d), which shows the error in the flow field reconstructed using only the recovered large-scale modes related to the entire flow field.

\section{Conclusion}\label{Sec:Conclusion}
In this paper, a Lagrangian-Eulerian multiscale data assimilation (LEMDA) framework is developed to facilitate the state estimation of multiscale flow fields. Despite the nonlinearity, the DA solutions can be written down using closed analytic formulae, which allows for efficiently reaching accurate solutions and avoiding empirical tunings. The Lagrangian and Eulerian components have their significant features. Lagrangian DA can be directly applied when particle trajectories are used as observations. Effective reduced-order filters can be derived to further enhance computational efficiency. The Eulerian DA scheme is systematically derived from the Boltzmann equation to advance a helpful and efficient method that recovers large-scale features. The Eulerian DA is also particularly valuable when particles have collisions, such as using sea ice floes as particles. In such a case, the direct Lagrangian DA is not the most appropriate choice for state estimation. The Eulerian and Lagrangian DAs are combined for multiscale DA. Parallel computing can be naturally incorporated to significantly accelerate computational efficiency.

There are several natural extensions of the current study, which remain as future work. First, only the momentum equations are used in the Eulerian DA. When the particles are represented by ice floes and other floats with geometric structures, angular momentum is another widely used measurement that helps state estimation. The governing equation of the angular momentum can be derived following the procedure in Section \ref{Subsec:Eulerian_Model}. Then, such an additional observational quantity can be incorporated into the DA framework to potentially help enhance the DA skill. Second, most of the numerical experiments in this paper consider a perfect model setup. It is essential to study the DA skill when model errors appear in the forecast model. Some crucial sources of model errors are statistical approximate errors using the linear stochastic model to approximate highly nonlinear true underlying systems and using linear drags to approximate the quadratic drag law. Another interesting topic is the optimal selection of a few Lagrangian trajectories to supplement the Eulerian observations that maximize the uncertainty reduction in the estimated flow field from the hybrid Eulerian-Lagrangian observations. Information criteria \cite{chen2023launching} are useful cost functions to resolve such an optimization problem. Furthermore, this work assumes the particles appear almost everywhere in the flow field. In practice, clouds may intervene in the observations (e.g., from satellite images) and bring about missing observations in a certain region. Therefore, studying the skill of LEMDA with partial observations will be a practically meaningful topic. Note that the DA considered in this work assumes continuous-in-time observations. It is natural to develop an analog using discrete-in-time observations. Finally, despite the advantage of exact analytic formulae in solving the posterior distributions in the framework developed here, the multiscale DA procedure can also be combined with the general ensemble DA approaches in many other applications.

\section*{Acknowledgments}
The research of S.N.S and N.C. was partially funded by the Office of Naval Research (ONR) Multidisciplinary University Research Initiative (MURI) award N00014-19-1-2421. J.H. is supported as a postdoc research associated under this grant. The research of Q.D. is partially supported by the Australian National Computing Infrastructure (NCI) national facility under grant zv32.
Data and codes are available at: \hyperlink{https://github.com/QuanlingDeng/LEMDA}{https://github.com/QuanlingDeng/LEMDA}.

\section{Supporting Information}

\subsection{Derivation of the reduced-order Lagrangian DA}
Recall the governing equation of the Lagrangian DA:
\begin{subequations}\label{Lagrangian_Tracer_Model_SI}
\begin{align}
  \frac{\d { \bfs{x}_l}}{\d t} &= \bfs{v}_l + \sigma_x\dot{\bfs{W}}_l, \label{Lagrangian_Tracer_Model_x_SI}\\
  \frac{\d \bfs{v}_l}{\d t} &= \beta(\bfs{u}-\bfs{v}_l),\label{Lagrangian_Tracer_Model_v_SI} \qquad\bfs{u}(\bfs{x},t) = \sum_{\bfs{k}\in\mathcal{K}}\hat{u}_{\bfs{k}}(t)e^{i\bfs{k}\bfs{x}}\bfs{r}_{\bfs{k}},\\
  \frac{\d\hat{u}_{\bfs{k}}}{\d t} &= - d_{\bfs{k}}  \hat{u}_{\bfs{k}} + f_{\bfs{k}}(t) + \sigma_{\bfs{k}}\dot{W}_{\bfs{k}},\label{Lagrangian_Flow_Model_SI}
\end{align}
\end{subequations}
where the subscript $\alpha$ is omitted for notation simplicity. The flow field can be regarded as having only one type of mode, such as the geophysically balanced mode that corresponds to the incompressible flow. This is also a natural choice to satisfy the condition that the tracers are (nearly) uniformly distributed. For this reason, the phase term $i\omega_{\bfs{k}}$ is omitted as well for simplicity. A similar process can be carried out for the derivation when $\omega_{\bfs{k}}$ is nonzero. The forcing $\bfs{f}_l$ in \eqref{Lagrangian_Tracer_Model_v_SI}, describing the contact force between different tracers, is also dropped to allow the flow field to dominate the tracer motion. Writing the equation \eqref{Lagrangian_Tracer_Model_SI} in the general form of \eqref{eq:filter}. The observed state $\bfs{Z}$ contains the displacements of all the tracers $\bfs{x}_l$, for $l=1,\ldots,2L$. The unobserved state $\bfs{U}$ involves the $2L$ velocity components of the tracers $\bfs{v}_l$ and the $|\mathcal{K}|$ spectral modes of the underlying flow field $\hat{u}_{\bfs{k}}$, where $|\mathcal{K}|$ is the total number of the spectral modes in the set $\mathcal{K}$. The observed and unobserved variables have dimensions $2L\times 1$ and $(2L+|\mathcal{K}|)\times 1$, respectively.

Recall the posterior covariance equation of the filtering solution:
\begin{equation}\label{eq:filter_R_SI}
\frac{\d\bfs{R}}{\d t} = \bfs{\Lambda}\bfs{R} + \bfs{R}\bfs{\Lambda}^\ast + \bfs{\Sigma}_\bfs{U}\bfs{\Sigma}_\bfs{U}^\ast - \bfs{R}\bfs{A}^\ast(\bfs{\Sigma}_\bfs{Z}\bfs{\Sigma}_\bfs{Z}^\ast)^{-1}\bfs{A}\bfs{R}.
\end{equation}
The reduced filter aims to simplify the evolution equation of the covariance $\bfs{R}$ in \eqref{eq:filter_R_SI}. Notably, the matrix $\bfs{R}$ can be written in the block form:
\begin{equation}\label{R_Block_SI}
  \bfs{R} = \left(
                 \begin{array}{cc}
                   \bfs{R}_{11} & \bfs{R}_{12} \\
                   \bfs{R}_{21} & \bfs{R}_{22} \\
                 \end{array}
               \right)
\end{equation}
where $\bfs{R}_{11}$ corresponds to the variance of the tracer velocity and $\bfs{R}_{22}$ contains the variance of the spectral modes of the underlying flow field. The cross-covariance block $\bfs{R}_{12}=\bfs{R}_{21}^*$ contains the coupling between the tracer velocity of the underlying flow field. The dimensions of $\bfs{R}_{11}$, $\bfs{R}_{22}$ and $\bfs{R}_{12}$ are $2L\times 2L$, $|\mathcal{K}|\times|\mathcal{K}|$, and $2L\times|\mathcal{K}|$, respectively. The matrix $\bfs{R}_{11}$ is full because the tracers nearby are highly correlated.  The matrix $\bfs{R}_{12}$ is also full. The off-diagonal components pass the information from the inferred tracer velocity based on their displacement to recover the underlying flow field in DA. If such a matrix becomes diagonal, then the underlying system loses observability to estimate the flow field from observations. The main reduction in the covariance lies in the block $\bfs{R}_{22}$. Note that all the spectral modes are coupled to form the flow velocity, and therefore, $\bfs{R}_{22}$ from DA is a full matrix. Nevertheless, since different spectral modes are, by design, independent of each other, it is expected that, with $L\to\infty$, the posterior covariance block $\bfs{R}_{22}$ also becomes a diagonal matrix. This has been shown when the velocities of the tracers and the underlying flow field are synchronized \cite{chen2014information}, assuming the flow field is incompressible. Furthermore, the block $\bfs{R}_{22}$ will asymptotically become a constant matrix as $L\to\infty$ and can be predetermined, which avoids running any governing equation. Deriving such a constant diagonal matrix is the primary goal in this subsection. The only assumption utilized here is that the tracers are distributed uniformly in the domain. This assumption is roughly satisfied when the underlying flow field is (nearly) incompressible and the drag coefficient $\beta$ is not too small.

The matrices in \eqref{eq:filter_R_SI} have the following form:
\begin{equation}\label{matrices_1_SI}
\begin{gathered}
  \bfs{\Lambda} = \left(
                           \begin{array}{cc}
                             -\beta\bfs{I}_{2L} & \bfs{Q} \\
                             \bfs{0} & -\bfs{D} \\
                           \end{array}
                         \right),\qquad \bfs{A} = \left(
                           \begin{array}{cc}
                             \bfs{I}_{2L} & \bfs{0} \\
                           \end{array}
                         \right),\\ \bfs{\Sigma}_\bfs{U}\bfs{\Sigma}_\bfs{U}^\ast = \left(
                           \begin{array}{cc}
                             \bfs{0} & \bfs{0} \\
                             \bfs{0} & \bfs{\Sigma}_\bfs{k}^2  \\
                           \end{array}\right),\qquad \bfs{\Sigma}_\bfs{Z}\bfs{\Sigma}_\bfs{Z}^\ast = \sigma_x^{2}\bfs{I}_{2L},
\end{gathered}
\end{equation}
where $\bfs{I}_{2L}$ is an identity matrix of size $2L\times2L$ and $\bfs{D}$ is a diagonal matrix of size $|\mathcal{K}|\times|\mathcal{K}|$ with the diagonal entries being $d_{\bfs{k}}$. The block matrix $\bfs{Q}$ is of size $2L\times|\mathcal{K}|$ with its entries being the two components of $e^{i\bfs{k}\bfs{x}_l}\bfs{r}_\bfs{k}$. The matrix $\bfs{\Sigma}_\bfs{k}^2$ is diagonal with size $|\mathcal{K}|\times|\mathcal{K}|$, where each diagonal component is $\sigma_{\bfs{k}}^2$.

In addition to $\bfs{\Sigma}_\bfs{U}\bfs{\Sigma}_\bfs{U}^\ast$, the remaining terms on the right-hand side of \eqref{eq:filter_R_SI} can be computed:
\begin{equation}\label{R_eqn_rhs1_SI}
\begin{split}
  \bfs{R}\bfs{A}^\ast(\bfs{\Sigma}_\bfs{Z}\bfs{\Sigma}_\bfs{Z}^\ast)^{-1}\bfs{A}\bfs{R} &= \left(
                           \begin{array}{cc}
                             \bfs{R}_{11} & \bfs{R}_{12} \\
                             \bfs{R}_{21} & \bfs{R}_{22} \\
                           \end{array}
                         \right)\left(
                           \begin{array}{c}
                             \bfs{I}_{2L} \\
                             \bfs{0} \\
                           \end{array}
                         \right)\sigma_{x}^{-2}\bfs{I}_{2L}\left(
                           \begin{array}{cc}
                             \bfs{I}_{2L} & \bfs{0} \\
                           \end{array}
                         \right)\left(
                           \begin{array}{cc}
                             \bfs{R}_{11} & \bfs{R}_{12} \\
                             \bfs{R}_{21} & \bfs{R}_{22} \\
                           \end{array}
                         \right)\\
                         &=\sigma_{x}^{-2}\left(
                           \begin{array}{cc}
                             \bfs{R}_{11}\bfs{R}_{11} & \bfs{R}_{11}\bfs{R}_{12} \\
                             \bfs{R}_{21}\bfs{R}_{11} & \bfs{R}_{21}\bfs{R}_{12} \\
                           \end{array}
                         \right),
\end{split}
\end{equation}
and
\begin{equation}\label{R_eqn_rhs2_SI}
\begin{split}
\bfs{\Lambda}\bfs{R} + \bfs{R}\bfs{\Lambda}^\ast &= \left(
                           \begin{array}{cc}
                             -\beta\bfs{I}_{2L} & \bfs{Q} \\
                             \bfs{0} & -\bfs{D} \\
                           \end{array}
                         \right)\left(
                           \begin{array}{cc}
                             \bfs{R}_{11} & \bfs{R}_{12} \\
                             \bfs{R}_{21} & \bfs{R}_{22} \\
                           \end{array}
                         \right) + \left(
                           \begin{array}{cc}
                             \bfs{R}_{11} & \bfs{R}_{12} \\
                             \bfs{R}_{21} & \bfs{R}_{22} \\
                           \end{array}
                         \right)\left(
                           \begin{array}{cc}
                             -\beta\bfs{I}_{2L} & \bfs{0} \\
                             \bfs{Q}^* & -\bfs{D} \\
                           \end{array}
                         \right)\\
                         &=\left(
                           \begin{array}{cc}
                             -\beta\bfs{R}_{11} + \bfs{Q}\bfs{R}_{21} & -\beta\bfs{R}_{12} + \bfs{Q}\bfs{R}_{22} \\
                             -\bfs{D}\bfs{R}_{21} & -\bfs{D}\bfs{R}_{22} \\
                           \end{array}
                         \right) + \left(
                           \begin{array}{cc}
                             -\beta\bfs{R}_{11} + \bfs{R}_{12}\bfs{Q}^* & -\bfs{R}_{12}\bfs{D} \\
                             -\beta\bfs{R}_{21} + \bfs{R}_{22}\bfs{Q}^* & -\bfs{R}_{22}\bfs{D} \\
                           \end{array}
                         \right).
\end{split}
\end{equation}
In light of \eqref{matrices_1_SI}, \eqref{R_eqn_rhs1_SI} and \eqref{R_eqn_rhs2_SI}, the time evolution of the three blocks in the posterior covariance $\bfs{R}$ can be written as
\begin{subequations}\label{Covariance_Evolution_SI}
\begin{align}
  \frac{\d \bfs{R}_{11}}{\d t} &= -2\beta\bfs{R}_{11} + \bfs{Q}\bfs{R}_{21} + \bfs{R}_{12}\bfs{Q}^* - \sigma_x^{-2}\bfs{R}_{11}\bfs{R}_{11},\label{Covariance_Evolution1_SI}\\
  \frac{\d \bfs{R}_{21}}{\d t} &= -\bfs{D}\bfs{R}_{21} -\beta\bfs{R}_{21} +\bfs{R}_{22}\bfs{Q}^*-\sigma_x^{-2}\bfs{R}_{21}\bfs{R}_{11},\label{Covariance_Evolution2_SI}\\
  \frac{\d \bfs{R}_{22}}{\d t} &= -2\bfs{D}\bfs{R}_{22} + \bfs{\Sigma}_{\bfs{k}}^2 - \sigma_x^{-2}\bfs{R}_{21}\bfs{R}_{12}.\label{Covariance_Evolution3_SI}
\end{align}
\end{subequations}
As time evolves, the distribution of the tracers will roughly have a steady behaviour. Although the exact distribution still varies in time due to the randomness in the flow field, the system can be regarded as reaching a quasi-equilibrium state. Under such a condition, the goal is to look for a solution that approximately gives
\begin{equation}\label{R_equilibrium_SI}
  \frac{\d \bfs{R}}{\d t }=0.
\end{equation}
With \eqref{R_equilibrium_SI} in hand, it can be deduced from \eqref{Covariance_Evolution2_SI} that
\begin{equation}\label{R21_SI}
\bfs{R}_{21}\approx \mathcal{A}\bfs{R}_{22}\bfs{Q}^*,
\end{equation}
where $\mathcal{A}$ is a diagonal matrix with its $i$-th diagonal entry corresponding to the $\bfs{k}$-th spectral mode being
\begin{equation}\label{alpha_SI}
\xi = \frac{1}{d_\bfs{k}+\beta+\sigma_x^{-2}\gamma}.
\end{equation}
Here, $\gamma$ is used to parameterize the effect of $\bfs{R}_{11}$, when it is multiplied to $\bfs{R}_{21}$. Note that similar to $d_\bfs{k}$, the two scalars $\xi$ and $\gamma$ depend on the mode $\bfs{k}$. But for notation simplicity, the subscript $\bfs{k}$ is omitted for these two scalars. Accordingly, the $i$-th diagonal component of $\bfs{R}_{22}$ in \eqref{Covariance_Evolution3_SI} becomes
\begin{equation}\label{Rk_SI}
\frac{\d R_{\bfs{k}}}{\d t}= -2d_\bfs{k}R_\bfs{k} +\sigma_\bfs{k}^2 - \sigma_x^{-2}\xi^2 L R_{k}^2,
\end{equation}
where \eqref{R21_SI} is used to replace $\bfs{R}_{21}$ in \eqref{Covariance_Evolution3_SI} such that \eqref{Rk_SI} becomes a closed equation of $R_{\bfs{k}}$, one of the diagonal components of $\bfs{R}_{22}$, itself. From \eqref{R21_SI} to \eqref{Rk_SI}, the mean-field theory is applied such that $\bfs{Q}^*\bfs{Q}\approx L\cdot\bfs{I}_{|\mathcal{K}|}$. This can be seen by noting that the $\bfs{m}$ and $\bfs{n}$ entry of $\bfs{Q}^*\bfs{Q}$ can be explicitly written as
\begin{equation}\label{matrixP_SI}
  \bfs{Q}^*\bfs{Q}_{\bfs{m},\bfs{n}} = \sum_{l=1}^L\exp(i(\bfs{m}-\bfs{n})\bfs{x}_l) \bfs{r}^*_{\bfs{n}}\bfs{r}_{\bfs{m}}.
\end{equation}
If the tracers are uniformly distributed, then the entries of $\bfs{P}_{\bfs{m},\bfs{n}}$ with $\bfs{m}\neq\bfs{n}$ become zero in the asymptotic limit according to the mean-field theory while those with $\bfs{m}=\bfs{n}$ equal to $L$.

The positive root of $R_{\bfs{k}}$ in \eqref{Rk_SI} is given by
\begin{equation}\label{root_Rk_SI}
  R_{\bfs{k}} = \frac{-d_k + \sqrt{d_k^2+\sigma_x^{-2}\xi^2L\sigma_k^2}}{\sigma_x^{-2}\xi^2L},
\end{equation}
which is the constant posterior variance associated with the component $\hat{u}_\bfs{k}$.

What remains is to determine $\xi$. Multiplying $\bfs{Q}$ by both sides of \eqref{R21_SI} yields
\begin{equation}\label{QR21_SI}
\bfs{Q}\bfs{R}_{21}\approx \mathcal{A}\bfs{Q}\bfs{R}_{22}\bfs{Q}^*.
\end{equation}
Thus each diagonal component of $\bfs{Q}\bfs{R}_{21}$ is approximated by $\xi L R_{\bfs{k}}$. Note that $R_{11}$ is not a diagonal matrix but only the diagonal entries are used here to determine $\xi$. Plugging the above result into \eqref{Covariance_Evolution1_SI} leads to the following equation for each diagonal component of $\bfs{R}_{11}$, namely the parameterized scalar $\gamma$,
\begin{equation}\label{R11_eqn_SI}
  -2\beta \gamma + 2\xi L \bfs{R}_{k}-\sigma_x^{-2}\gamma^2=0,
\end{equation}
the solution of which is given by
\begin{equation}\label{solution_gamma_SI}
  \gamma = \frac{-\beta+\sqrt{\beta^2+2\sigma_x^{-2}\xi L R_{\bfs{k}}}}{\sigma_x^{-2}}.
\end{equation}
The diagonal entry $R_{\bfs{k}}$, namely the posterior variance of mode $\hat{u}_{\bfs{k}}$, is then obtained by solving the three-dimensional coupled algebraic equations \eqref{alpha_SI}, \eqref{root_Rk_SI} and \eqref{solution_gamma_SI}. These three equations have three unknowns $R_{\bfs{k}}$, $\xi$ and $\gamma$. A standard Newton's iterative method is utilized to find the solution.

Once all the diagonal entries $R_{\bfs{k}}$ of $\bfs{R}_{22}$ are pre-determined, only the equations \eqref{Covariance_Evolution1_SI} and \eqref{Covariance_Evolution2_SI} need to be solved. Then finally, the constant diagonal covariance block for the flow modes can be predetermined by the smoother solution. The computational time and storage for the smoothing solutions can similarly be reduced.

\subsection{Discretization scheme of the Eulerian equations}

In general, the equations \eqref{eq:numdensity} and \eqref{eq:mv} are of the following form of conservation law
\begin{equation} \label{eq:conlaw}
    \frac{\partial w }{\partial t} + \nabla \cdot \big(\bfs{v} w \big) = f(\bfs{x}, t)
\end{equation}
with an initial solution $w_0.$

This section describes the spatiotemporal discretization scheme implemented in the Eulerian DA. Let $(h_x, h_y)$ be the spatial grid spacing and $dt$ be the time step. The spatial and temporal derivatives are approximated using finite differences. In the DA implementation, the first-order upwind finite difference scheme, finite volume method, and center-finite difference scheme for the discretization of the term $\nabla \cdot \big(\bfs{v} w \big)$ are tested. They all yield similar results (based on the RMSE and pattern correlation skill scores). As has been pointed out in past studies \cite{grote2006stable,majda2007explicit}, it is possible for an unstable discretization (such as the standard centered scheme) to be stabilized through the assimilation of observational data. Below, the centered finite difference discretization of $\nabla \cdot \big(\bfs{v} w \big)$ is utilized for its symmetry and simplicity and good results.

The temporal discretization can be performed using an explicit time-stepping scheme, such as the forward Euler method:
\begin{equation}
\frac{\partial w}{\partial t} \approx \frac{w_{j,i}^{n+1} - w_{j,i}^n}{dt},
\end{equation}
where $w_{j,i}^n$ is the value of $w$ at spatial grid point $(x_i, y_j)$ and time step $n$.
The spatial discretization is using a central difference scheme for the flux term:
\begin{equation}
\nabla \cdot \big(\bfs{v} w \big) \approx \frac{(\bfs{v} w)_{j,i+1}^n - (\bfs{v} w)_{j,i-1}^n }{2h_x} + \frac{(\bfs{v} w)_{j+1,i}^n - (\bfs{v} w)_{j-1,i}^n }{2h_y},
\end{equation}
where $(j,i)$ represents the indices in $y$ and $x$ dimensions following the Matlab matrix representation.

Substituting the spatial and temporal discretizations into \eqref{eq:conlaw} yields the following scheme:
\begin{equation}
\frac{w_{j,i}^{n+1} - w_{j,i}^n}{dt} +
\nabla \cdot \big(\bfs{v} w \big) \approx \frac{w_{j,i}^{n+1} - w_{j,i}^n}{dt} +
\frac{(\bfs{v} w)_{j,i+1}^n - (\bfs{v} w)_{j,i-1}^n }{2h_x} + \frac{(\bfs{v} w)_{j+1,i}^n - (\bfs{v} w)_{j-1,i}^n }{2h_y} = f_{j,i}^n.
\end{equation}
Lastly, the boundary terms are updated using periodic boundary conditions.

\subsection{Estimating the random noise coefficient in the processes of Eulerian observations} \label{sec:eudanoise}
The processes of the statistical moment equations associated with the Eulerian observations, e.g., the momentum equations \eqref{eq:mv}, are a set of partial differential equations. However, the actual Eulerian observations are the averaged values of these quantities at coarse mesh grids. Therefore, the observational processes in \eqref{CGNS_X} are the spatial discrete form of \eqref{eq:mv} and other statistical moment equations. At least two sources of errors need to be considered in carrying out the Eulerian DA. One is the discretization error, and another is the error due to using a finite number of particles in computing the observed statistics. These potential errors are characterized by the random noise in \eqref{CGNS_X}. Since the noise describes the uncertainty and influences the DA solution, it is essential to provide an effective estimation of the random noise coefficient $\bfs{\Sigma}_\bfs{Z}$, which is written as $\bfs{\Sigma}_\bfs{Z}=\bfs{\Sigma}_{\bfs{Z},dis} + \bfs{\Sigma}_{\bfs{Z},spl}$, representing the uncertainty associated with the discretization and the finite number of samples, respectively. For simplicity, assume both $\bfs{\Sigma}_{\bfs{Z},dis}$ and $\bfs{\Sigma}_{\bfs{Z},spl}$ are diagonal matrices, implying the independence of the noise at different grid cells.

Consider the spatial discretized equation \eqref{eq:mv} evaluated at a specific location $\bfs{x}^*$:
\begin{subequations}\label{Comparison_Discretization}
\begin{align}
  \left.\frac{\d \langle\rho\bfs{v}\rangle}{\d t}\right|_{\bfs{x}^*} &= G(\rho\bfs{v}),\label{Comparison_Discretization_fine}\\
  \left.\frac{\d \langle\widetilde{\rho\bfs{v}}\rangle}{\d t}\right|_{\bfs{x}^*} &= \widetilde{G}(\widetilde{\rho\bfs{v}}),\label{Comparison_Discretization_coarse}
\end{align}
\end{subequations}
where a fine discretization is used in \eqref{Comparison_Discretization_fine} while the spatial discretization of \eqref{Comparison_Discretization_coarse} is based on a coarse mesh as in LEMDA. The solution in \eqref{Comparison_Discretization_fine} can be regarded as the truth when studying the discretization error in \eqref{Comparison_Discretization_coarse}.
Taking the difference between these two equations yields
\begin{equation}\label{Difference_Discretization}
  \frac{\d \zeta_{\bfs{x}^*}}{\d t} := \left.\frac{\d \left(\langle\rho\bfs{v}\rangle- \langle\widetilde{\rho\bfs{v}}\rangle\right)}{\d t}\right|_{\bfs{x}^*} =  G(\rho\bfs{v})-\widetilde{G}(\widetilde{\rho\bfs{v}}).
\end{equation}
Note that since $\langle\rho\bfs{v}\rangle$ is computed using a much finer solver, a spatial average over the multiple nearby locations within the same grid cell as $\langle\widetilde{\rho\bfs{v}}\rangle$ needs to be taken in computing \eqref{Difference_Discretization}. The equation \eqref{Difference_Discretization} is run forward by only one time step from $t$ to $t+\Delta{t}$ with the initial condition $\zeta_{\bfs{x}^*}(t)=0$. Therefore, the error is solely due to the difference between the functions $G$ and $\widetilde{G}$ that accounts for the discretization error. By computing the one-step solution of $\zeta_{\bfs{x}^*}((j+1)\Delta{t})$ with the initial condition of $\langle\rho\bfs{v}\rangle(j\Delta{t})$ being taken at different time instants for $j = 1, 2, \ldots, J$, the resulting $J$ points of $\zeta_{\bfs{x}^*}$ are then used to estimate one diagonal component of the noise coefficient $\bfs{\Sigma}_{\bfs{Z},dis}$, which is the standard deviation of these $\zeta_{\bfs{x}^*}$.

Next, the diagonal component of $\bfs{\Sigma}_{\bfs{Z},spl}$ is estimated in light of the standard Monte Carlo error analysis. Denote by $n$ the number of particles average over time in one specific grid cell. This gives the corresponding diagonal component of $\bfs{\Sigma}_{\bfs{Z},spl}=\sigma_x/\sqrt{n}$.

A practical simplification of $\bfs{\Sigma}_{\bfs{Z},dis}$ and $\bfs{\Sigma}_{\bfs{Z},spl}$ is to assume both of them are constant diagonal matrices, which is appropriate if the underlying flow field is statistically homogeneous. Recall that $L$ is the total number of particles in the domain. Denote by $N=N_x\times N_y$ the number of the discrete grid cells, where $N_x$ and $N_y$ are the number of grids in the $x$ and the $y$ directions, respectively. Denote by $\zeta$ the set that contains all the $NJ$ points from the $\zeta_{\bfs{x}^*}$ at different $\bfs{x}^*$. Similarly, the average number of particles in each grid cell is $L/N$. The constant diagonal entries of $\bfs{\Sigma}_{\bfs{Z},dis}$ and $\bfs{\Sigma}_{\bfs{Z},spl}$ are thus given by $\mbox{std}(\zeta)$ and $\sigma_xN/L$, respectively. These will be the estimated noise coefficients $\bfs{\Sigma}_{\bfs{Z}}$ used in the numerical experiments. In some cases, especially in the presence of model error, noise inflation can be further added based on the calculated values from the above procedure.

\subsection{Interpolating missing values at Eulerian grid points}\label{Subsec:Interpolation}

It is not always guaranteed that each grid cell contains at least one Lagrangian particle throughout time, especially when the total number of tracers is limited or the underlying flow field is strongly compressible. To fill in the missing data, a standard thin-plate spline interpolation method is applied \cite{duchon1977splines}.

\begin{figure}[ht]
    \includegraphics[width=1.0\textwidth]{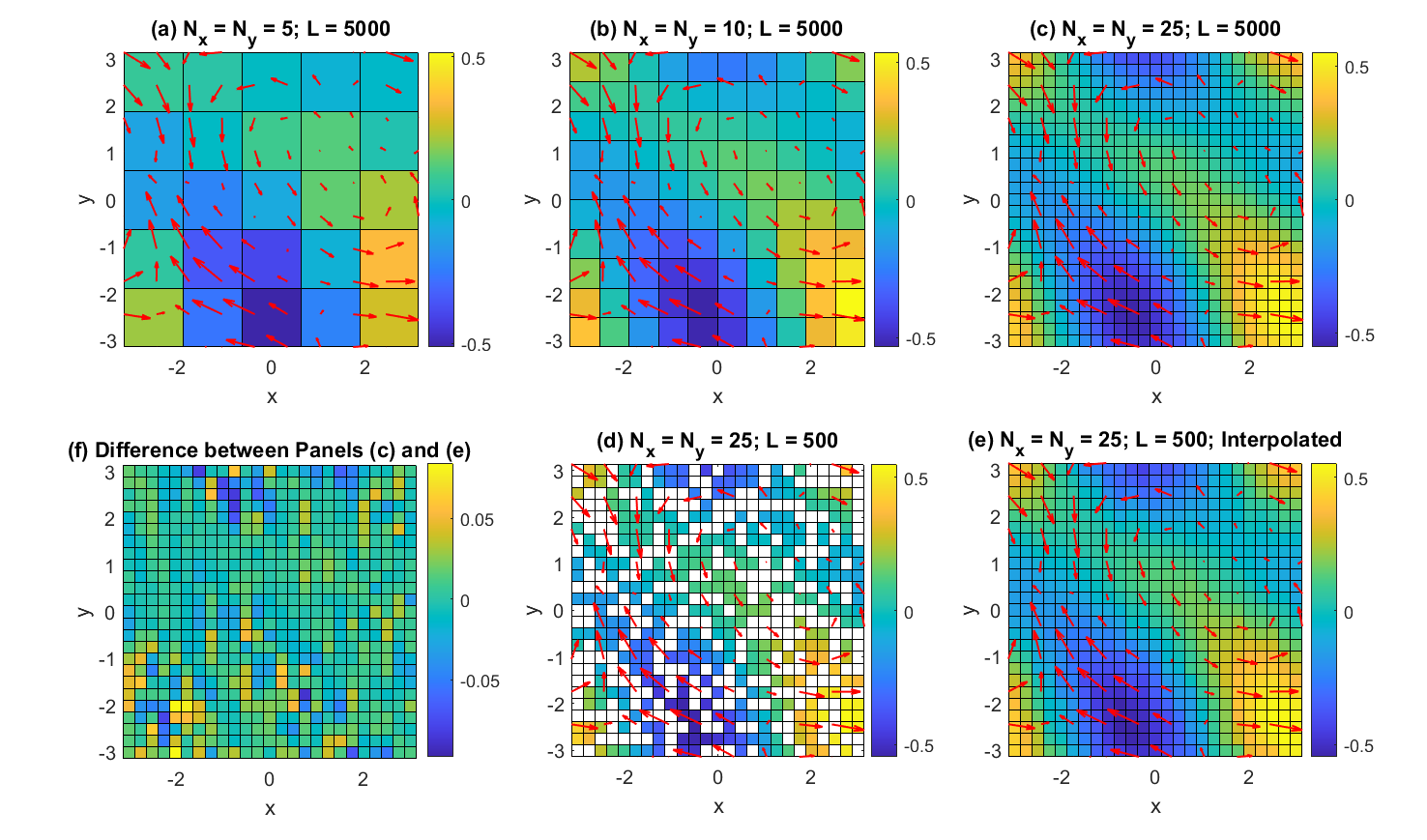}
    \caption{A proof-of-concept validation. Panels (a)--(c) include the spatial distribution of the zonal momentum at a specific time $t=2$ with different mesh grids using $L=5000$ particles. Panel (d) shows the spatial distribution of the zonal momentum with $N_x=N_y=25$ but using only $L=500$ Lagrangian particles. Panel (e) shows the result after applying the spatial interpolation to Panel (d). Panel (f) shows the difference between Panel (d) and Panel (e).
    }
    \label{fig:SI_Interpolation}
\end{figure}

A numerical simulation is shown in Figure \ref{fig:SI_Interpolation} as a proof-of-concept validation. In this simulation, an incompressible flow field with $K_{\mbox{max}}=1$ is adopted such that there are a total of $9$ Fourier modes in the random flow field. It is represented using the red arrows in the figure. The contour plot in Panel (a) shows the spatial distribution of the zonal momentum (e.g., the first component of $\langle\rho \bfs{v} \rangle$ in \eqref{eq:mv}) with three different spatial discretizations: $N_x=N_y = 5,10$ and $25$. First, the Eulerian observations are computed based on $L=5000$ particles. Panels (a)--(c) include the spatial distribution of the zonal momentum at a specific time $t=2$ with different mesh grids. Notably, with a spatial resolution of $N_x=N_y=25$, there are, on average, only $8$ particles in each grid cell. Despite having a small number of particles, the zonal momentum has a smooth transition in space. Panel (d) shows the spatial distribution of the zonal momentum with $N_x=N_y=25$ but using only $L=500$ Lagrangian particles. As expected, several grid cells contain no particles. Then, the spatial interpolation is applied, and the result is shown in Panel (e). The spatial distribution after the interpolation is similar to the case using $L=5000$ particles. Their difference, as is shown in Panel (f), is negligible.

\bibliographystyle{plain}
\bibliography{references}

\begin{thebibliography}{10}

\bibitem{apte2008data}
A~Apte, Christopher~KRT Jones, AM~Stuart, and Jochen Voss.
\newblock Data assimilation: {M}athematical and statistical perspectives.
\newblock {\em International journal for numerical methods in fluids},
  56(8):1033--1046, 2008.

\bibitem{apte2013impact}
A~Apte and CKRT Jones.
\newblock The impact of nonlinearity in {L}agrangian data assimilation.
\newblock {\em Nonlinear Processes in Geophysics}, 20(3):329--341, 2013.

\bibitem{apte2008bayesian}
Amit Apte, Christopher~KRT Jones, and AM~Stuart.
\newblock A {B}ayesian approach to {L}agrangian data assimilation.
\newblock {\em Tellus A: Dynamic Meteorology and Oceanography}, 60(2):336--347,
  2008.

\bibitem{asch2016data}
Mark Asch, Marc Bocquet, and Ma{\"e}lle Nodet.
\newblock {\em Data assimilation: methods, algorithms, and applications}.
\newblock SIAM, 2016.

\bibitem{azouani2014continuous}
Abderrahim Azouani, Eric Olson, and Edriss~S Titi.
\newblock Continuous data assimilation using general interpolant observables.
\newblock {\em Journal of Nonlinear Science}, 24:277--304, 2014.

\bibitem{bach2023filtering}
Eviatar Bach, Tim Colonius, and Andrew Stuart.
\newblock Filtering dynamical systems using observations of statistics.
\newblock {\em arXiv preprint arXiv:2308.05484}, 2023.

\bibitem{bergemann2012ensemble}
Kay Bergemann and Sebastian Reich.
\newblock An ensemble {K}alman-{B}ucy filter for continuous data assimilation.
\newblock {\em Meteorologische Zeitschrift}, 21(3):213, 2012.

\bibitem{berner2017stochastic}
Judith Berner, Ulrich Achatz, Lauriane Batte, Lisa Bengtsson, Alvaro
  De~La~Camara, Hannah~M Christensen, Matteo Colangeli, Danielle~RB Coleman,
  Daan Crommelin, Stamen~I Dolaptchiev, et~al.
\newblock Stochastic parameterization: {T}oward a new view of weather and
  climate models.
\newblock {\em Bulletin of the American Meteorological Society},
  98(3):565--588, 2017.

\bibitem{blunden2019look}
J~Blunden and DS~Arndt.
\newblock A look at 2018: {T}akeaway points from the state of the climate
  supplement.
\newblock {\em Bulletin of the American Meteorological Society},
  100(9):1625--1636, 2019.

\bibitem{branicki2013non}
Michal Branicki, Nan Chen, and Andrew~J Majda.
\newblock Non-{G}aussian test models for prediction and state estimation with
  model errors.
\newblock {\em Chinese Annals of Mathematics, Series B}, 34(1):29--64, 2013.

\bibitem{branicki2018accuracy}
Michal Branicki, Andrew~J Majda, and Kody~JH Law.
\newblock Accuracy of some approximate {G}aussian filters for the
  {N}avier--{S}tokes equation in the presence of model error.
\newblock {\em Multiscale Modeling \& Simulation}, 16(4):1756--1794, 2018.

\bibitem{brocker2010variational}
Jochen Br{\"o}cker.
\newblock On variational data assimilation in continuous time.
\newblock {\em Quarterly Journal of the Royal Meteorological Society},
  136(652):1906--1919, 2010.

\bibitem{bucy2005filtering}
Richard~S Bucy and Peter~D Joseph.
\newblock {\em Filtering for stochastic processes with applications to
  guidance}, volume 326.
\newblock American Mathematical Soc., 2005.

\bibitem{businger1996balloons}
Steven Businger, Steven~R Chiswell, Warren~C Ulmer, and Randy Johnson.
\newblock Balloons as a {L}agrangian measurement platform for atmospheric
  research.
\newblock {\em Journal of Geophysical Research: Atmospheres},
  101(D2):4363--4376, 1996.

\bibitem{carlson2021sensitivity}
Elizabeth Carlson and Adam Larios.
\newblock Sensitivity analysis for the 2d {N}avier--{S}tokes equations with
  applications to continuous data assimilation.
\newblock {\em Journal of Nonlinear Science}, 31(5):84, 2021.

\bibitem{castellari2001prediction}
Sergio Castellari, Annalisa Griffa, Tamay~M {\"O}zg{\"o}kmen, and Pierre-Marie
  Poulain.
\newblock Prediction of particle trajectories in the {A}driatic sea using
  {L}agrangian data assimilation.
\newblock {\em Journal of Marine Systems}, 29(1-4):33--50, 2001.

\bibitem{centurioni2017global}
Luca Centurioni, Andr{\'a}s Hor{\'a}nyi, Carla Cardinali, Etienne Charpentier,
  and Rick Lumpkin.
\newblock A global ocean observing system for measuring sea level atmospheric
  pressure: {E}ffects and impacts on numerical weather prediction.
\newblock {\em Bulletin of the American Meteorological Society},
  98(2):231--238, 2017.

\bibitem{cercignani1988boltzmann}
Carlo Cercignani and Carlo Cercignani.
\newblock {\em The Boltzmann equation}.
\newblock Springer, 1988.

\bibitem{chaikin1995principles}
Paul~M Chaikin, Tom~C Lubensky, and Thomas~A Witten.
\newblock {\em Principles of condensed matter physics}, volume~10.
\newblock Cambridge university press Cambridge, 1995.

\bibitem{chen2020learning}
Nan Chen.
\newblock Learning nonlinear turbulent dynamics from partial observations via
  analytically solvable conditional statistics.
\newblock {\em Journal of Computational Physics}, 418:109635, 2020.

\bibitem{chen2023stochastic}
Nan Chen.
\newblock {\em Stochastic Methods for Modeling and Predicting Complex Dynamical
  Systems: Uncertainty Quantification, State Estimation, and Reduced-Order
  Models}.
\newblock Springer Nature, 2023.

\bibitem{chen2023uncertainty}
Nan Chen and Shubin Fu.
\newblock Uncertainty quantification of nonlinear {L}agrangian data
  assimilation using linear stochastic forecast models.
\newblock {\em Physica D: Nonlinear Phenomena}, page 133784, 2023.

\bibitem{chen2021lagrangian}
Nan Chen, Shubin Fu, and Georgy Manucharyan.
\newblock Lagrangian data assimilation and parameter estimation of an idealized
  sea ice discrete element model.
\newblock {\em Journal of Advances in Modeling Earth Systems},
  13(10):e2021MS002513, 2021.

\bibitem{chen2022efficient}
Nan Chen, Shubin Fu, and Georgy~E Manucharyan.
\newblock An efficient and statistically accurate {L}agrangian data
  assimilation algorithm with applications to discrete element sea ice models.
\newblock {\em Journal of Computational Physics}, 455:111000, 2022.

\bibitem{chen2022conditional}
Nan Chen, Yingda Li, and Honghu Liu.
\newblock Conditional {G}aussian nonlinear system: {A} fast preconditioner and
  a cheap surrogate model for complex nonlinear systems.
\newblock {\em Chaos: An Interdisciplinary Journal of Nonlinear Science},
  32(5):053122, 2022.

\bibitem{chen2023launching}
Nan Chen, Evelyn Lunasin, and Stephen Wiggins.
\newblock Launching drifter observations in the presence of uncertainty.
\newblock {\em arXiv preprint arXiv:2307.12779}, 2023.

\bibitem{chen2016model}
Nan Chen and Andrew~J Majda.
\newblock Model error in filtering random compressible flows utilizing noisy
  {L}agrangian tracers.
\newblock {\em Monthly Weather Review}, 144(11):4037--4061, 2016.

\bibitem{chen2018conditional}
Nan Chen and Andrew~J Majda.
\newblock Conditional {G}aussian systems for multiscale nonlinear stochastic
  systems: {P}rediction, state estimation and uncertainty quantification.
\newblock {\em Entropy}, 20(7):509, 2018.

\bibitem{chen2014information}
Nan Chen, Andrew~J Majda, and Xin~T Tong.
\newblock Information barriers for noisy {L}agrangian tracers in filtering
  random incompressible flows.
\newblock {\em Nonlinearity}, 27(9):2133, 2014.

\bibitem{chen2015noisy}
Nan Chen, Andrew~J Majda, and Xin~T Tong.
\newblock Noisy {L}agrangian tracers for filtering random rotating compressible
  flows.
\newblock {\em Journal of Nonlinear Science}, 25(3):451--488, 2015.

\bibitem{covington2022bridging}
Jeffrey Covington, Nan Chen, and Monica~M Wilhelmus.
\newblock Bridging gaps in the climate observation network: {A} physics-based
  nonlinear dynamical interpolation of {L}agrangian ice floe measurements via
  data-driven stochastic models.
\newblock {\em Journal of Advances in Modeling Earth Systems},
  14(9):2022MS003218, 2022.

\bibitem{cundall1979discrete}
Peter~A Cundall and Otto~DL Strack.
\newblock A discrete numerical model for granular assemblies.
\newblock {\em geotechnique}, 29(1):47--65, 1979.

\bibitem{damsgaard2018application}
Anders Damsgaard, Alistair Adcroft, and Olga Sergienko.
\newblock Application of discrete element methods to approximate sea ice
  dynamics.
\newblock {\em Journal of Advances in Modeling Earth Systems},
  10(9):2228--2244, 2018.

\bibitem{delsole2005predictability}
Timothy DelSole.
\newblock Predictability and information theory. {P}art {II}: {I}mperfect
  forecasts.
\newblock {\em Journal of the atmospheric sciences}, 62(9):3368--3381, 2005.

\bibitem{deng2023particle}
Quanling Deng, Samuel~N. Stechmann, and Nan Chen.
\newblock Particle-continuum multiscale modeling of sea ice floes.
\newblock {\em Multiscale Modeling \& Simulation}, 22(1):230--255, 2024.

\bibitem{desamsetti2019efficient}
Srinivas Desamsetti, Hari~Prasad Dasari, Sabique Langodan, Edriss~S Titi, Omar
  Knio, and Ibrahim Hoteit.
\newblock Efficient dynamical downscaling of general circulation models using
  continuous data assimilation.
\newblock {\em Quarterly Journal of the Royal Meteorological Society},
  145(724):3175--3194, 2019.

\bibitem{duchon1977splines}
Jean Duchon.
\newblock Splines minimizing rotation-invariant semi-norms in {S}obolev spaces.
\newblock In {\em Constructive Theory of Functions of Several Variables:
  Proceedings of a Conference Held at Oberwolfach April 25--May 1, 1976}, pages
  85--100. Springer, 1977.

\bibitem{einicke2012smoothing}
Garry Einicke.
\newblock {\em Smoothing, filtering and prediction: Estimating the past,
  present and future}.
\newblock BoD--Books on Demand, 2012.

\bibitem{evensen2000ensemble}
Geir Evensen and Peter~Jan Van~Leeuwen.
\newblock An ensemble {K}alman smoother for nonlinear dynamics.
\newblock {\em Monthly Weather Review}, 128(6):1852--1867, 2000.

\bibitem{farchi2021using}
Alban Farchi, Patrick Laloyaux, Massimo Bonavita, and Marc Bocquet.
\newblock Using machine learning to correct model error in data assimilation
  and forecast applications.
\newblock {\em Quarterly Journal of the Royal Meteorological Society},
  147(739):3067--3084, 2021.

\bibitem{farhat2020data}
Aseel Farhat, Nathan~E Glatt-Holtz, Vincent~R Martinez, Shane~A McQuarrie, and
  Jared~P Whitehead.
\newblock Data assimilation in large prandtl {R}ayleigh--{B}enard convection
  from thermal measurements.
\newblock {\em SIAM Journal on Applied Dynamical Systems}, 19(1):510--540,
  2020.

\bibitem{farrell1993stochastic}
Brian~F Farrell and Petros~J Ioannou.
\newblock Stochastic forcing of the linearized {N}avier--{S}tokes equations.
\newblock {\em Physics of Fluids A: Fluid Dynamics}, 5(11):2600--2609, 1993.

\bibitem{garcia2022structured}
Guillermo Garc{\'\i}a-S{\'a}nchez, Ana~M Mancho, Antonio~G Ramos, Josep Coca,
  and Stephen Wiggins.
\newblock Structured pathways in the turbulence organizing recent oil spill
  events in the eastern mediterranean.
\newblock {\em Scientific Reports}, 12(1):3662, 2022.

\bibitem{gardiner1985handbook}
Crispin~W Gardiner et~al.
\newblock {\em Handbook of stochastic methods}, volume~3.
\newblock springer Berlin, 1985.

\bibitem{giannakis2012quantifying}
Dimitrios Giannakis and Andrew~J Majda.
\newblock Quantifying the predictive skill in long-range forecasting. {P}art
  {I}: Coarse-grained predictions in a simple ocean model.
\newblock {\em Journal of climate}, 25(6):1793--1813, 2012.

\bibitem{gould2004argo}
John Gould, Dean Roemmich, Susan Wijffels, Howard Freeland, Mark Ignaszewsky,
  Xu~Jianping, Sylvie Pouliquen, Yves Desaubies, Uwe Send, Kopillil
  Radhakrishnan, et~al.
\newblock Argo profiling floats bring new era of in situ ocean observations.
\newblock {\em Eos, Transactions American Geophysical Union}, 85(19):185--191,
  2004.

\bibitem{griffa2007lagrangian}
Annalisa Griffa, AD~Kirwan~Jr, Arthur~J Mariano, Tamay {\"O}zg{\"o}kmen, and
  H~Thomas Rossby.
\newblock {\em Lagrangian analysis and prediction of coastal and ocean
  dynamics}.
\newblock Cambridge University Press, 2007.

\bibitem{grote2006stable}
Marcus~J Grote and Andrew~J Majda.
\newblock Stable time filtering of strongly unstable spatially extended
  systems.
\newblock {\em Proceedings of the National Academy of Sciences},
  103(20):7548--7553, 2006.

\bibitem{hamill2005accounting}
Thomas~M Hamill and Jeffrey~S Whitaker.
\newblock Accounting for the error due to unresolved scales in ensemble data
  assimilation: A comparison of different approaches.
\newblock {\em Monthly weather review}, 133(11):3132--3147, 2005.

\bibitem{harlim2008filtering}
J~Harlim and AJ~Majda.
\newblock Filtering nonlinear dynamical systems with linear stochastic models.
\newblock {\em Nonlinearity}, 21(6):1281, 2008.

\bibitem{harlim2013test}
John Harlim and Andrew~J Majda.
\newblock Test models for filtering and prediction of moisture-coupled tropical
  waves.
\newblock {\em Quarterly Journal of the Royal Meteorological Society},
  139(670):119--136, 2013.

\bibitem{harris2004introduction}
Stewart Harris.
\newblock {\em An introduction to the theory of the Boltzmann equation}.
\newblock Courier Corporation, 2004.

\bibitem{honnorat2009lagrangian}
Marc Honnorat, J{\'e}r{\^o}me Monnier, and Fran{\c{c}}ois-Xavier Le~Dimet.
\newblock Lagrangian data assimilation for river hydraulics simulations.
\newblock {\em Computing and visualization in science}, 12(5):235--246, 2009.

\bibitem{hyndman2006another}
Rob~J Hyndman and Anne~B Koehler.
\newblock Another look at measures of forecast accuracy.
\newblock {\em International journal of forecasting}, 22(4):679--688, 2006.

\bibitem{ide2002lagrangian}
Kayo Ide, Leonid Kuznetsov, and Christopher~KRT Jones.
\newblock Lagrangian data assimilation for point vortex systems.
\newblock {\em Journal of Turbulence}, 3(1):053, 2002.

\bibitem{jazwinski2007stochastic}
Andrew~H Jazwinski.
\newblock {\em Stochastic processes and filtering theory}.
\newblock Courier Corporation, 2007.

\bibitem{kalman1961new}
Rudolph~E Kalman and Richard~S Bucy.
\newblock New results in linear filtering and prediction theory.
\newblock {\em Journal of Fluids Engineering}, 83:95--108, 1961.

\bibitem{kalnay2003atmospheric}
Eugenia Kalnay.
\newblock {\em Atmospheric modeling, data assimilation and predictability}.
\newblock Cambridge university press, 2003.

\bibitem{kalnay1996ncep}
Eugenia Kalnay, Masao Kanamitsu, Robert Kistler, William Collins, Dennis
  Deaven, Lev Gandin, Mark Iredell, Suranjana Saha, Glenn White, John Woollen,
  et~al.
\newblock The {NCEP}/{NCAR} 40-year reanalysis project.
\newblock {\em Bulletin of the American meteorological Society},
  77(3):437--472, 1996.

\bibitem{kang2012filtering}
Emily~L Kang and John Harlim.
\newblock Filtering nonlinear spatio-temporal chaos with autoregressive linear
  stochastic models.
\newblock {\em Physica D: Nonlinear Phenomena}, 241(12):1099--1113, 2012.

\bibitem{kleeman2002measuring}
Richard Kleeman.
\newblock Measuring dynamical prediction utility using relative entropy.
\newblock {\em Journal of the atmospheric sciences}, 59(13):2057--2072, 2002.

\bibitem{kleeman2011information}
Richard Kleeman.
\newblock Information theory and dynamical system predictability.
\newblock {\em Entropy}, 13(3):612--649, 2011.

\bibitem{kullback1987letter}
Solomon Kullback.
\newblock Letter to the editor: The {K}ullback-{L}eibler distance.
\newblock {\em AMERICAN STATISTICIAN}, 1987.

\bibitem{kullback1951information}
Solomon Kullback and Richard~A Leibler.
\newblock On information and sufficiency.
\newblock {\em The annals of mathematical statistics}, 22(1):79--86, 1951.

\bibitem{lahoz2010data}
Boris Khattatov~William Lahoz and Richard Menard.
\newblock {\em Data assimilation}.
\newblock Springer, 2010.

\bibitem{law2015data}
Kody Law, Andrew Stuart, and Kostas Zygalakis.
\newblock Data assimilation.
\newblock {\em Cham, Switzerland: Springer}, 214, 2015.

\bibitem{lean2021continuous}
P~Lean, EV~H{\'o}lm, M~Bonavita, N~Bormann, AP~McNally, and Heikki
  J{\"a}rvinen.
\newblock Continuous data assimilation for global numerical weather prediction.
\newblock {\em Quarterly Journal of the Royal Meteorological Society},
  147(734):273--288, 2021.

\bibitem{li2020predictability}
Ying Li and Samuel~N Stechmann.
\newblock Predictability of tropical rainfall and waves: {E}stimates from
  observational data.
\newblock {\em Quarterly Journal of the Royal Meteorological Society},
  146(729):1668--1684, 2020.

\bibitem{liptser2013statistics}
Robert~S Liptser and Albert~N Shiryaev.
\newblock {\em Statistics of random processes {II}: {A}pplications}, volume~6.
\newblock Springer Science \& Business Media, 2013.

\bibitem{lopez2023level}
Rigoberto~Moncada Lopez, Mukund Gupta, Andrew Thompson, and Jose Andrade.
\newblock Level set discrete element method for modeling sea ice floes.
\newblock {\em Computer Methods in Applied Mechanics and Engineering},
  406:115891, 2023.

\bibitem{majda2003introduction}
Andrew Majda.
\newblock {\em Introduction to PDEs and Waves for the Atmosphere and Ocean},
  volume~9.
\newblock American Mathematical Soc., 2003.

\bibitem{majda2005information}
Andrew Majda, Rafail~V Abramov, and Marcus~J Grote.
\newblock {\em Information theory and stochastics for multiscale nonlinear
  systems}, volume~25.
\newblock American Mathematical Soc., 2005.

\bibitem{majda2016introduction}
Andrew~J Majda.
\newblock {\em Introduction to turbulent dynamical systems in complex systems}.
\newblock Springer, 2016.

\bibitem{majda2018model}
Andrew~J Majda and Nan Chen.
\newblock Model error, information barriers, state estimation and prediction in
  complex multiscale systems.
\newblock {\em Entropy}, 20(9):644, 2018.

\bibitem{majda2007explicit}
Andrew~J Majda and Marcus~J Grote.
\newblock Explicit off-line criteria for stable accurate time filtering of
  strongly unstable spatially extended systems.
\newblock {\em Proceedings of the National Academy of Sciences},
  104(4):1124--1129, 2007.

\bibitem{majda2012filtering}
Andrew~J Majda and John Harlim.
\newblock {\em Filtering complex turbulent systems}.
\newblock Cambridge University Press, 2012.

\bibitem{manucharyan2022spinning}
Georgy~E Manucharyan, Rosalinda Lopez-Acosta, and Monica~M Wilhelmus.
\newblock Spinning ice floes reveal intensification of mesoscale eddies in the
  western arctic ocean.
\newblock {\em Scientific Reports}, 12(1):7070, 2022.

\bibitem{manucharyan2022subzero}
Georgy~E Manucharyan and Brandon~P Montemuro.
\newblock {SubZero}: {A} sea ice model with an explicit representation of the
  floe life cycle.
\newblock {\em Journal of Advances in Modeling Earth Systems},
  14(12):e2022MS003247, 2022.

\bibitem{mu2018arctic}
Longjiang Mu, Martin Losch, Qinghua Yang, Robert Ricker, Svetlana~N Losa, and
  Lars Nerger.
\newblock Arctic-wide sea ice thickness estimates from combining satellite
  remote sensing data and a dynamic ice-ocean model with data assimilation
  during the {CryoSat}-2 period.
\newblock {\em Journal of Geophysical Research: Oceans}, 123(11):7763--7780,
  2018.

\bibitem{potyondy2004bonded}
David~O Potyondy and PA~Cundall.
\newblock A bonded-particle model for rock.
\newblock {\em International journal of rock mechanics and mining sciences},
  41(8):1329--1364, 2004.

\bibitem{rebholz2021simple}
Leo~G Rebholz and Camille Zerfas.
\newblock Simple and efficient continuous data assimilation of evolution
  equations via algebraic nudging.
\newblock {\em Numerical Methods for Partial Differential Equations},
  37(3):2588--2612, 2021.

\bibitem{salman2008using}
H~Salman, K~Ide, and Christopher~KRT Jones.
\newblock Using flow geometry for drifter deployment in {L}agrangian data
  assimilation.
\newblock {\em Tellus A: Dynamic Meteorology and Oceanography}, 60(2):321--335,
  2008.

\bibitem{sarkka2023bayesian}
Simo S{\"a}rkk{\"a} and Lennart Svensson.
\newblock {\em Bayesian filtering and smoothing}, volume~17.
\newblock Cambridge university press, 2023.

\bibitem{silva2021new}
Thiago~MD Silva, Sinesio Pesco, Abelardo Barreto~Jr, and Mustafa Onur.
\newblock A new procedure for generating data covariance inflation factors for
  ensemble smoother with multiple data assimilation.
\newblock {\em Computers \& Geosciences}, 150:104722, 2021.

\bibitem{sun2019lagrangian}
Luyu Sun and Stephen~G Penny.
\newblock Lagrangian data assimilation of surface drifters in a double-gyre
  ocean model using the local ensemble transform {K}alman filter.
\newblock {\em Monthly Weather Review}, 147(12):4533--4551, 2019.

\bibitem{taylor2023submesoscale}
John~R Taylor and Andrew~F Thompson.
\newblock Submesoscale dynamics in the upper ocean.
\newblock {\em Annual Review of Fluid Mechanics}, 55:103--127, 2023.

\bibitem{uppala2005era}
Sakari~M Uppala, PW~K{\aa}llberg, Adrian~J Simmons, U~Andrae, V~Da~Costa
  Bechtold, M~Fiorino, JK~Gibson, J~Haseler, A~Hernandez, GA~Kelly, et~al.
\newblock The {ERA}-40 re-analysis.
\newblock {\em Quarterly Journal of the Royal Meteorological Society: A journal
  of the atmospheric sciences, applied meteorology and physical oceanography},
  131(612):2961--3012, 2005.

\bibitem{van2012origin}
Erik Van~Sebille, Matthew~H England, and Gary Froyland.
\newblock Origin, dynamics and evolution of ocean garbage patches from observed
  surface drifters.
\newblock {\em Environmental Research Letters}, 7(4):044040, 2012.

\bibitem{yang2013feedback}
Tao Yang, Prashant~G Mehta, and Sean~P Meyn.
\newblock Feedback particle filter.
\newblock {\em IEEE transactions on Automatic control}, 58(10):2465--2480,
  2013.

\end{thebibliography}

\end{document}